%% file: main.tex
\documentclass[ALICE,manyauthors]{cernphprep}
\usepackage[comma,square,numbers,sort&compress]{natbib}
\usepackage[T1]{fontenc}
\usepackage{hyperref}
\usepackage{lineno}
\usepackage{xcolor}
\usepackage{xspace}
\usepackage{multirow}
\usepackage{makecell}
\usepackage{booktabs}
\usepackage{upgreek}
\usepackage{comment}
\usepackage{amsmath}
\usepackage{orcidlink}
\begin{document}
\input{commands.tex}

\begin{titlepage}
\PHyear{2023}       
\PHnumber{159}      
\PHdate{02 August}  


\title{Study of flavor dependence of the baryon-to-meson ratio \\ in proton--proton collisions at $\pmb{\s = 13~\TeV}$}
\ShortTitle{Flavor dependence of the baryon-to-meson ratio in pp at $\s = 13~\TeV$}   
\Collaboration{ALICE Collaboration\thanks{See Appendix~\ref{app:collab} for the list of collaboration members}}
\ShortAuthor{ALICE Collaboration} 

\begin{abstract}
The production cross sections of \Dzero and \Lc hadrons originating from beauty-hadron decays (i.e.~non-prompt) were measured for the first time at midrapidity ($|y|<0.5$) by the ALICE Collaboration in proton--proton collisions at a center-of-mass energy $\s=13~\TeV$. They are described within uncertainties by perturbative QCD calculations employing the fragmentation fractions of beauty quarks to baryons measured at forward rapidity by the LHCb Collaboration. The \bbbar production cross section per unit of rapidity at midrapidity, estimated from these measurements, is $\de\sigma_{\bbbar}/\de y|_{|y|<0.5} = 83.1 \pm 3.5 (\mathrm{stat}) \pm 5.4(\mathrm{syst}) ^{+12.3}_{-3.2} (\mathrm{extrap})~\mub$. The baryon-to-meson ratios are computed to investigate the hadronization mechanism of beauty quarks. The non-prompt $\Lc/\Dzero$ production ratio has a similar trend to the one measured for the promptly produced charmed particles and to the p$/\pip$ and $\lmb/\kzero$ ratios, suggesting a similar baryon-formation mechanism among light, strange, charm, and beauty hadrons. The \pt-integrated non-prompt $\Lc/\Dzero$ ratio is found to be significantly higher than the one measured in \ee collisions.

\end{abstract}
\end{titlepage}

\setcounter{page}{2} 


\input{intro}

\input{dataanalysis}
\input{results}


\newenvironment{acknowledgement}{\relax}{\relax}
\begin{acknowledgement}
\section*{Acknowledgements}
\input{fa_2023-06-23_Opt_C.tex}
\end{acknowledgement}

\bibliographystyle{utphys}   
\bibliography{bibliography}

\newpage
\appendix

%
%

\section{The ALICE Collaboration}
\label{app:collab}
\input{2023-06-23-Alice_Authorlist_2023-06-23_Opt_C.tex}  
\end{document}

%% file: commands.tex
%

\newcommand{\pp}           {pp\xspace}
\newcommand{\ppbar}        {\mbox{$\mathrm {p\overline{p}}$}\xspace}
\newcommand{\XeXe}         {\mbox{Xe--Xe}\xspace}
\newcommand{\PbPb}         {\mbox{Pb--Pb}\xspace}
\newcommand{\pA}           {\mbox{pA}\xspace}
\newcommand{\pPb}          {\mbox{p--Pb}\xspace}
\newcommand{\AuAu}         {\mbox{Au--Au}\xspace}
\newcommand{\dAu}          {\mbox{d--Au}\xspace}

\newcommand{\s}            {\ensuremath{\sqrt{s}}\xspace}
\newcommand{\snn}          {\ensuremath{\sqrt{s_{\mathrm{NN}}}}\xspace}
\newcommand{\pt}           {\ensuremath{p_{\rm T}}\xspace}
\newcommand{\meanpt}       {$\langle p_{\mathrm{T}}\rangle$\xspace}
\newcommand{\ycms}         {\ensuremath{y_{\rm CMS}}\xspace}
\newcommand{\ylab}         {\ensuremath{y_{\rm lab}}\xspace}
\newcommand{\etarange}[1]  {\mbox{$\left | \eta \right |~<~#1$}}
\newcommand{\yrange}[1]    {\mbox{$\left | y \right |~<~#1$}}
\newcommand{\dndy}         {\ensuremath{\mathrm{d}N_\mathrm{ch}/\mathrm{d}y}\xspace}
\newcommand{\dndeta}       {\ensuremath{\mathrm{d}N_\mathrm{ch}/\mathrm{d}\eta}\xspace}
\newcommand{\avdndeta}     {\ensuremath{\langle\dndeta\rangle}\xspace}
\newcommand{\dNdy}         {\ensuremath{\mathrm{d}N_\mathrm{ch}/\mathrm{d}y}\xspace}
\newcommand{\Npart}        {\ensuremath{N_\mathrm{part}}\xspace}
\newcommand{\Ncoll}        {\ensuremath{N_\mathrm{coll}}\xspace}
\newcommand{\dEdx}         {\ensuremath{\textrm{d}E/\textrm{d}x}\xspace}
\newcommand{\RpPb}         {\ensuremath{R_{\rm pPb}}\xspace}
\newcommand{\RAA}          {\ensuremath{R_{\rm AA}}\xspace}
\newcommand{\QpPb}         {\ensuremath{Q_{\rm pPb}}\xspace}
\newcommand{\Qcp}          {\ensuremath{Q_{\rm CP}}\xspace}
\newcommand{\Ntrkl}        {\ensuremath{N_\mathrm{trkl}}\xspace}
\newcommand{\averNtrkl}    {\ensuremath{\langle N_\mathrm{trkl}\rangle}\xspace}
\newcommand{\Nch}          {\ensuremath{N_\mathrm{ch}}\xspace}
\newcommand{\averNch}      {\ensuremath{\langle N_\mathrm{ch}\rangle}\xspace}
\newcommand{\zVtx}         {\ensuremath{z_\mathrm{vtx}}\xspace}
\newcommand*\code[1]{\texttt{#1}}
\newcommand{\fprompt}      {\ensuremath{f_\mathrm{prompt}}\xspace}
\newcommand{\fnonprompt}   {\ensuremath{f_\mathrm{non\text{-}prompt}}\xspace}
\newcommand{\AccEff}       {\ensuremath{\mbox{$\text{Acc}\times\epsilon$}}\xspace}
\newcommand{\effNP}[1]         {\ensuremath{(\mathrm{Acc}\times\epsilon)^\mathrm{non\text{-}prompt}_{#1}}\xspace}
\newcommand{\effP}[1]          {\ensuremath{(\mathrm{Acc}\times\epsilon)^\mathrm{prompt}_\mathrm{#1}}\xspace}
\newcommand{\AccEffNP}         {\ensuremath{(\mathrm{Acc}\times\epsilon)_{\mathrm{non\text{-}prompt}}}\xspace}
\newcommand{\rawY}[1]          {\ensuremath{Y_\mathrm{#1}}\xspace}
\newcommand{\Nraw}               {\ensuremath{N_\mathrm{raw}}\xspace}
\newcommand{\fP}           {\ensuremath{f_\mathrm{p}}\xspace}
\newcommand{\fNP}          {\ensuremath{f_\mathrm{np}}\xspace}
\newcommand{\Np}               {\ensuremath{N_\mathrm{prompt}}\xspace}
\newcommand{\Nnp}              {\ensuremath{N_\mathrm{non\text{-}prompt}}\xspace}
\newcommand{\HbtoLc}       {\ensuremath{\mathrm{h_b}\to\Lc+\mathrm{X}}\xspace}
\newcommand{\HbtoHc}       {\ensuremath{\mathrm{h_b}\to\mathrm{h_c}+\mathrm{X}}\xspace}
\newcommand{\fbtoLb}       {\ensuremath{f(\mathrm{b}\to\Lb)}\xspace}
\newcommand{\fbtoHb}       {\ensuremath{f(\mathrm{b}\to\mathrm{h_b})}\xspace}
\newcommand{\fbtoB}       {\ensuremath{f(\mathrm{b}\to\mathrm{B})}\xspace}
\newcommand{\fbtoBee}      {\ensuremath{f(\mathrm{b}\to\mathrm{B})_{\mathrm{e^{+}e^{-}}}}\xspace}
\newcommand{\fbtoLbLHCb}       {\ensuremath{f(\mathrm{b}\to\Lb)_{\mathrm{LHCb}}\xspace}}

\newcommand{\LT}           {L{\'e}vy-Tsallis\xspace}
\newcommand{\TeV}          {\ensuremath{\mathrm{TeV}}\xspace}
\newcommand{\GeV}          {\ensuremath{\mathrm{GeV}}\xspace}
\newcommand{\MeV}          {\ensuremath{\mathrm{MeV}}\xspace}
\newcommand{\lumi}         {\ensuremath{\mathcal{L}}\xspace}
\newcommand{\de}           {\ensuremath{\mathrm{d}}\xspace}
\newcommand{\mub}          {\ensuremath{\upmu\mathrm{b}}\xspace}

\newcommand{\ITS}          {\rm{ITS}\xspace}
\newcommand{\TOF}          {\rm{TOF}\xspace}
\newcommand{\ZDC}          {\rm{ZDC}\xspace}
\newcommand{\ZDCs}         {\rm{ZDCs}\xspace}
\newcommand{\ZNA}          {\rm{ZNA}\xspace}
\newcommand{\ZNC}          {\rm{ZNC}\xspace}
\newcommand{\SPD}          {\rm{SPD}\xspace}
\newcommand{\SDD}          {\rm{SDD}\xspace}
\newcommand{\SSD}          {\rm{SSD}\xspace}
\newcommand{\TPC}          {\rm{TPC}\xspace}
\newcommand{\TRD}          {\rm{TRD}\xspace}
\newcommand{\VZERO}        {\rm{V0}\xspace}
\newcommand{\VZEROA}       {\rm{V0A}\xspace}
\newcommand{\VZEROC}       {\rm{V0C}\xspace}
\newcommand{\Vdecay} 	   {\ensuremath{V^{0}}\xspace}

\newcommand{\ee}                {\ensuremath{\mathrm{e^{+}e^{-}}}\xspace} 
\newcommand{\pip}               {\ensuremath{\uppi^{+}}\xspace}
\newcommand{\pim}               {\ensuremath{\uppi^{-}}\xspace}
\newcommand{\kap}               {\ensuremath{\rm{K}^{+}}\xspace}
\newcommand{\kam}               {\ensuremath{\rm{K}^{-}}\xspace}
\newcommand{\pbar}              {\ensuremath{\rm\overline{p}}\xspace}
\newcommand{\kzero}             {\ensuremath{{\rm K}^{0}_{\rm{S}}}\xspace}
\newcommand{\lmb}               {\ensuremath{\Lambda}\xspace}
\newcommand{\almb}              {\ensuremath{\overline{\Lambda}}\xspace}
\newcommand{\Om}                {\ensuremath{\Omega^-}\xspace}
\newcommand{\Mo}                {\ensuremath{\overline{\Omega}^+}\xspace}
\newcommand{\X}                 {\ensuremath{\Xi^-}\xspace}
\newcommand{\Ix}                {\ensuremath{\overline{\Xi}^+}\xspace}
\newcommand{\Xis}               {\ensuremath{\Xi^{\pm}}\xspace}
\newcommand{\Oms}               {\ensuremath{\Omega^{\pm}}\xspace}
\newcommand{\degree}            {\ensuremath{^{\rm o}}\xspace}
\newcommand{\DzerotoKpi}        {\ensuremath{{\rm D}^0 \to {\rm K}^-\uppi^+}\xspace}
\newcommand{\DplustoKpipi}      {\ensuremath{{\rm D}^+\to {\rm K}^-\uppi^+\uppi^+}\xspace}
\newcommand{\DstartoDpi}        {\ensuremath{{\rm D}^{*+} \to {\rm D}^0 \uppi^+}\xspace}
\newcommand{\DstartoDpiPrecise} {\ensuremath{{\rm D^{*+}(2010)\to D^0\uppi^+}}\xspace}
\newcommand{\Dstophipi}         {\ensuremath{{\rm D_s^{+}\to \upphi\uppi^+}}\xspace}
\newcommand{\DstophipitoKKpi}   {\ensuremath{{\rm D_s^{+}\to \upphi\uppi^+\to K^-K^+\uppi^+}}\xspace}
\newcommand{\phitoKK}           {\ensuremath{{\rm \upphi\to  K^-K^+}}\xspace}
\newcommand{\Dzero}             {\ensuremath{\mathrm{D^0}}\xspace}
\newcommand{\Dzerobar}          {\ensuremath{\mathrm{\overline{D}^0}}\xspace}
\newcommand{\Dstar}             {\ensuremath{\mathrm{D^{*+}}}\xspace}
\newcommand{\Dstarm}            {\ensuremath{\mathrm{D^{*-}}}\xspace}
\newcommand{\Dplus}             {\ensuremath{\mathrm{D^+}}\xspace}
\newcommand{\Dminus}            {\ensuremath{\mathrm{D^-}}\xspace}
\newcommand{\Ds}                {\ensuremath{\mathrm{D_s^+}}\xspace}
\newcommand{\Lc}                {\ensuremath{\Lambda_\mathrm{c}^+}\xspace}
\newcommand{\Lcpm}              {\ensuremath{\Lambda_\mathrm{c}^\pm}\xspace}
\newcommand{\Ks}                {\ensuremath{\mathrm{K_S^0}}\xspace}
\newcommand{\LctopKzeros}       {\ensuremath{\Lambda_\mathrm{c}^+\to\mathrm{p}\Ks}\xspace}
\newcommand{\Kstopipi}          {\ensuremath{\Ks\to\uppi^+\uppi^-}\xspace}
\newcommand{\LctopKpi}          {\ensuremath{\Lambda_\mathrm{c}^+\to\mathrm{pK^-\uppi^+}}\xspace}
\newcommand{\KKpi}              {\ensuremath{{\rm K^-K^+\uppi^+}}\xspace}
\newcommand{\Bplus}             {\ensuremath{\mathrm{B^+}}\xspace}
\newcommand{\Bs}                {\ensuremath{\mathrm{B_s^0}}\xspace}
\newcommand{\Bzero}             {\ensuremath{\mathrm{B^0}}\xspace}
\newcommand{\Lb}                {\ensuremath{\Lambda_\mathrm{b}^0}\xspace}
\newcommand{\Jpsi}                {\ensuremath{\mathrm{J}/\uppsi}\xspace}
\newcommand{\bbbar}             {\ensuremath{\mathrm{b\overline{b}}}\xspace}
\newcommand{\ccbar}             {\ensuremath{\mathrm{c\overline{c}}}\xspace}
\newcommand{\Hc}                {\ensuremath{\mathrm{h_c}}\xspace}
\newcommand{\Hb}                {\ensuremath{\mathrm{h_b}}\xspace}

\newcommand{\fonll}            {\textsc{fonll}\xspace}
\newcommand{\nlo}             {\textsc{nlo}\xspace}
\newcommand{\nnlo}             {\textsc{nnlo}\xspace}
\newcommand{\powheg}           {\textsc{Powheg}\xspace}
\newcommand{\gmvfns}           {\textsc{gm-vfns}\xspace}
\newcommand{\pythia}           {\textsc{Pythia~8}\xspace}
\newcommand{\pythiaprec}       {\textsc{Pythia~8.243}\xspace}
\newcommand{\geantthree}       {\textsc{Geant3}\xspace}
\newcommand{\clrblc}           {\textsc{clr-blc}\xspace}
\newcommand{\cteq}             {\textsc{cteq6.6}\xspace}
\newcommand{\ctnloten}             {\textsc{ct10nlo}\xspace}
\newcommand{\ctnlofourteen}             {\textsc{ct14nlo}\xspace}
\newcommand{\rqm}           {\textsc{rqm}\xspace}
\newcommand{\tamu}            {\textsc{tamu}\xspace}
\newcommand{\Lint}             {\ensuremath{\mathcal{L}_\mathrm{int}}\xspace}

%% file: intro.tex
\section{Introduction}
Measurements of open charm- and beauty-meson production in proton--proton (pp) collisions are successfully described by quantum chromodynamics (QCD) calculations based on the factorization of soft (non-perturbative) and hard (perturbative) processes~\cite{ALICE:2019nxm,ALICE:2021mgk,ALICE:2021edd,ALICE:2019nuy,CMS:2016plw,CMS:2017qjw,CMS:2018bwt,LHCb:2015swx,LHCb:2016ikn,LHCb:2017vec,ATLAS:2015igt,ALICE:2022wpn}. Within the collinear factorization approach, the production cross sections of heavy-flavor hadrons are computed as the convolution of (i) the parton distribution functions (PDFs) of the incoming protons, (ii) the perturbative partonic cross section, and (iii) the fragmentation functions (FFs) describing the transition from the heavy quark to the hadron. The partonic cross section in these calculations is typically computed at next-to-leading order accuracy with all-order resummation of next-to-leading logarithms (e.g., \fonll~\cite{Cacciari:1998it,Cacciari:2001td,Cacciari:2012ny} and \gmvfns~\cite{Kniehl:2004fy,Kniehl:2012ti,Benzke:2017yjn,Kramer:2017gct,Helenius:2018uul,Bolzoni:2013vya}), but recently calculations at next-to-next-to-leading order (\nnlo) also became available~\cite{Catani:2020kkl}. The FFs are instead parametrized from measurements performed in \ee or ep collisions~\cite{Braaten:1994bz}, assuming that the hadronization process of charm and beauty quarks is independent of the collision system.

Recent measurements of charm baryon-to-meson production ratios and fragmentation fractions (i.e. the probability of charm quark to fragment into a specific hadron) at midrapidity in pp collisions at LHC energies showed significant deviations from the values measured at \ee and ep colliders~\cite{ALICE:2021bli,ALICE:2020wfu,ALICE:2020wla,ALICE:2021rzj,ALICE:2022cop,CMS:2019uws,ALICE:2021dhb,ALICE:2022exq,ALICE:2021psx}, demonstrating that the assumption of universality of the hadronization process across the collision systems has to be reconsidered. As reported in Ref.~\cite{ALICE:2021dhb}, the observed baryon-to-meson enhancement also leads to an increase of the derived \ccbar production cross section at midrapidity compared to previous measurements, where prompt D-meson cross sections and fragmentation fractions from \ee were used. 
Models based on the hadronization via coalescence, i.e.~recombination of heavy and light quarks close in space and with similar velocity, are able to reproduce the magnitude as well as the \pt dependence of the measured $\Lc/\Dzero$ ratio~\cite{Song:2018tpv,Minissale:2020bif}. An alternative explanation is provided by statistical hadronization models, if an augmented set of yet unobserved charm-baryon states predicted by the relativistic-quark model~\cite{Ebert:2011kk} and lattice QCD~\cite{He:2019tik} is considered. Moreover, string fragmentation models including color-reconnection mechanisms beyond the leading-color (\clrblc) approximation introduce new topologies through the contributions of ``junctions” that fragment into baryons, thus providing an augmented baryon production~\cite{Christiansen:2015yqa}. In the beauty sector, the measurements of \Lb-baryon production relative to that of B mesons at forward rapidity by the LHCb Collaboration show a modification of the fragmentation fractions among collision systems similar to that observed for charm quarks at midrapidity~\cite{LHCb:2015qvk,LHCb:2019fns}. However, the rapidity-dependent baryon-to-meson enhancement with respect to values measured at \ee and ep colliders is still not fully understood and explored. Different enhancement was observed for the charm baryon-to-meson ratio between midrapidity and forward rapidity for $\pt < 8~\GeV/c$ and for different colliding systems (i.e. \pp, \pPb)~\cite{ALICE:2021rzj,LHCb:2013xam,LHCb:2018weo}. Beauty measurements at midrapidity are available only at high transverse momentum ($\pt > 7~\GeV/c$)~\cite{CMS:2017uoy,CMS:2018eso,CMS:2016plw,ATLAS:2013cia,CMS:2011oft,CMS:2011kew,CMS:2011pdu,CMS:2012wje}, hence a firm conclusion cannot be drawn.

In this article, the first measurements of the \pt-differential and \pt-integrated production cross sections of \Dzero mesons and \Lc baryons at midrapidity ($|y|<0.5$) originating from beauty-hadron decays (denoted as non-prompt) in pp collisions at a center-of-mass energy of $\s=13~\TeV$ are reported. The differential measurements performed in $1<\pt<24~\GeV/c$ and $2<\pt<24~\GeV/c$ for \Dzero and \Lc hadrons, respectively, are extrapolated down to $\pt=0$ to compute the \pt-integrated production cross sections and compute the \bbbar production cross section at midrapidity. Finally, the non-prompt $\Lc/\Dzero$ production ratio is compared with the production ratio of prompt $\Lc$ and $\Dzero$, with the p$/\pip$ and $\lmb/\kzero$ ratios, as well as with predictions based on the statistical hadronization model \tamu~\cite{He:2022tod} or the \pythia Monte Carlo (MC) generator~\cite{Sjostrand:2014zea} to investigate the hadronization mechanisms of beauty quarks into baryons and mesons.

%% file: dataanalysis.tex
\section{Experimental apparatus and data sample}
The description of the ALICE apparatus and of its performance is presented in detail in Refs.~\cite{ALICE:2014sbx,ALICE:2008ngc}.
The detectors used for the reconstruction of non-prompt charm-hadron decays are the Inner Tracking System (ITS), the Time Projection Chamber (TPC), and the Time-Of-Flight detector (TOF). These are located in the central barrel, which covers the pseudorapidity interval $|\eta|<0.9$ inside a solenoid magnet of field strength ${B = 0.5 \rm~T}$. The ITS and the TPC are used for tracking charged particles and reconstructing particle decay vertices. Particle identification (PID) is performed via the measurement of the specific energy loss ${\rm d}E/{\rm d}x$ in the TPC and of the flight time with the TOF detector. 

The data used for this analysis were collected during pp collisions at $\s = 13~\TeV$, with a minimum bias (MB) trigger. The latter requires coincident signals in the two scintillator arrays covering the intervals 2.8~$<\eta<$~5.1 (V0A) and -3.7~$<\eta<$~-1.7 (V0C). Offline selections were applied to remove background from beam--gas collisions, as described in Ref.~\cite{ALICE:2021rzj}. Only events with a primary vertex reconstructed within $\pm$10 cm from the nominal interaction point along the beam line were analyzed. Events with multiple primary vertices were rejected in order to remove collision pileup in the same bunch crossing. The remaining undetected pileup is negligible. The selected events correspond to an integrated luminosity $\mathcal{L}_{\rm int} = 31.9 \pm 0.5$\,${\rm nb}^{-1}$~\cite{ALICE:2021leo}.

\section{Data analysis}
\subsection{$\pmb{\Dzero}$ and $\pmb{\Lc}$ raw yields}
\label{sec:raw_yield}
The \Dzero meson, \Lc baryon, and their charge conjugates were reconstructed via the hadronic decay channels \DzerotoKpi (branching ratio $\mathrm{BR}=3.95\pm 0.03\%$), \LctopKzeros ($\mathrm{BR}=1.59\pm 0.08\%$, followed by $\rm K^{0}_{S} \rightarrow \rm \uppi^+\rm \uppi^-$ with a ${\rm BR}$ of $69.20 \pm 0.05\%$), and \LctopKpi ($\mathrm{BR}=6.28\pm 0.32\%$)~\cite{Workman:2022ynf}. The \Dzero-meson and \Lc-baryon candidates were reconstructed by combining pairs or triplets of tracks reconstructed with the proper charge, while for the \LctopKzeros channel the V-shaped decay of the \kzero meson into two pion-track candidates was combined with a proton-track candidate, using a Kalman-Filter vertexing algorithm~\cite{6550037}. All daughter tracks were required to be reconstructed within $|\eta|<$0.8 and to have at least 70 reconstructed space points in the TPC. For all decay products, at least one cluster was required in either of the two SPD layers.
As a consequence of these track-quality selections, the detector acceptance for the candidates falls steeply to zero for $|y|\approx0.5$ at \pt$\approx$~0 and at $|y|\approx0.8$ for $\pt > 5~\GeV/c$. Thus, a fiducial-acceptance selection ${|y|<y_{\rm fid}(\pt)}$ was applied, increasing from 0.5 at $\pt = 0$ to the maximum value of 0.8 at $\pt \geq 5~\GeV/c$.

A multi-class classification approach based on Boosted Decision Trees (BDT) algorithms provided by the \textsc{XGBoost}~\cite{Chen:2016XST,barioglio_luca_2021_5070132} library, was used to simultaneously reduce the combinatorial background, and to separate the prompt and non-prompt components for \Dzero mesons and \Lc baryons. Indeed, the inclusive sample of \Dzero and \Lc is dominated by the prompt hadrons and the non-prompt ones are a small fraction which can be enhanced exploiting the larger displacement due to the beauty-hadron decay topology. Signal samples of prompt and non-prompt candidates for the BDT training were obtained from MC simulations produced with the \pythiaprec~\cite{Sjostrand:2014zea} event generator with the Monash 2013 tune~\cite{Skands:2014pea}. The generated particles were propagated through the ALICE apparatus using the \geantthree transport code~\cite{Brun:1994aa}. Background samples were extracted from the candidate invariant-mass distributions inside the window of $5\sigma < |\Delta M| < 9\sigma$ in data, where $\Delta M$ is the difference between the candidate invariant mass and the \Dzero(\Lc) mass, and $\sigma$ is the invariant-mass resolution.  Loose selections on the decay kinematics, the decay topology, and the PID of the decay particles were applied to the candidates before the training. The BDTs were trained in each \pt interval using different variables related to the displaced decay-vertex topology and the PID of the decay tracks, similarly as described in Refs.~\cite{ALICE:2021mgk,ALICE:2021npz}. 
The BDT outputs are related to the candidate's probability to be either prompt or non-prompt \Dzero(\Lc) meson~(baryon) or combinatorial background. Selections on the BDT output probabilities of being background and non-prompt were optimized to obtain a high non-prompt \Dzero and \Lc fractions (i.e.~$\geq$~60$\%$ in each \pt interval) while maintaining a reliable signal extraction. The sample of candidates passing the BDT selection is denoted in the following as non-prompt enhanced sample. Based on the selections on the BDT outputs, samples dominated by non-prompt (prompt) candidates were selected by requiring low BDT probability for a candidate to be combinatorial background and a high BDT probability to be non-prompt (prompt).

\begin{figure}[!htb]
\begin{center}
\includegraphics[width=0.8\textwidth]{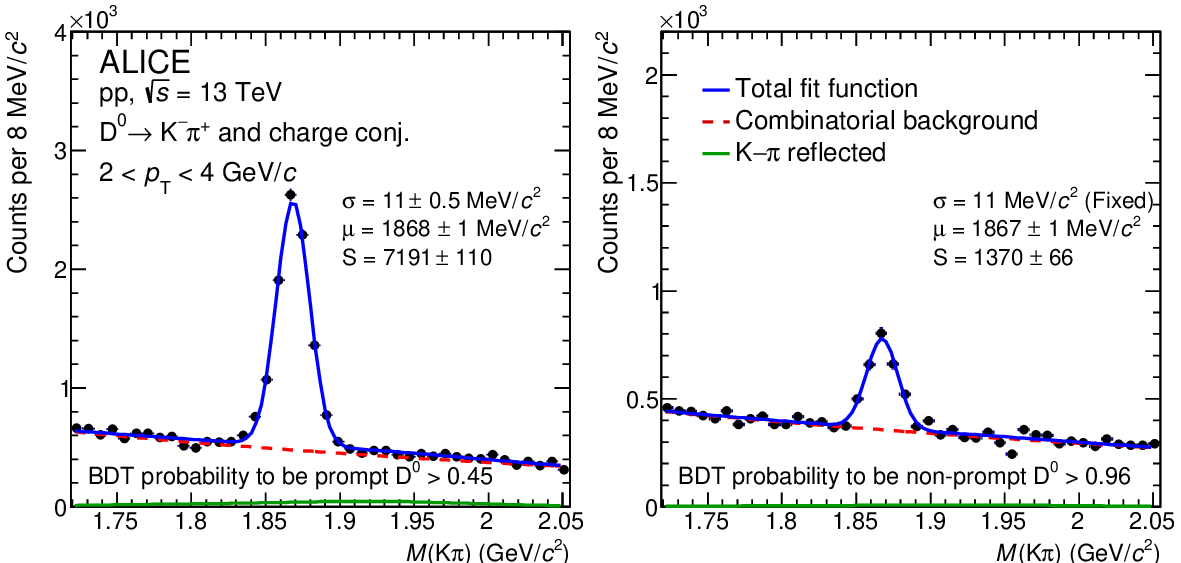}
\includegraphics[width=0.8\textwidth]{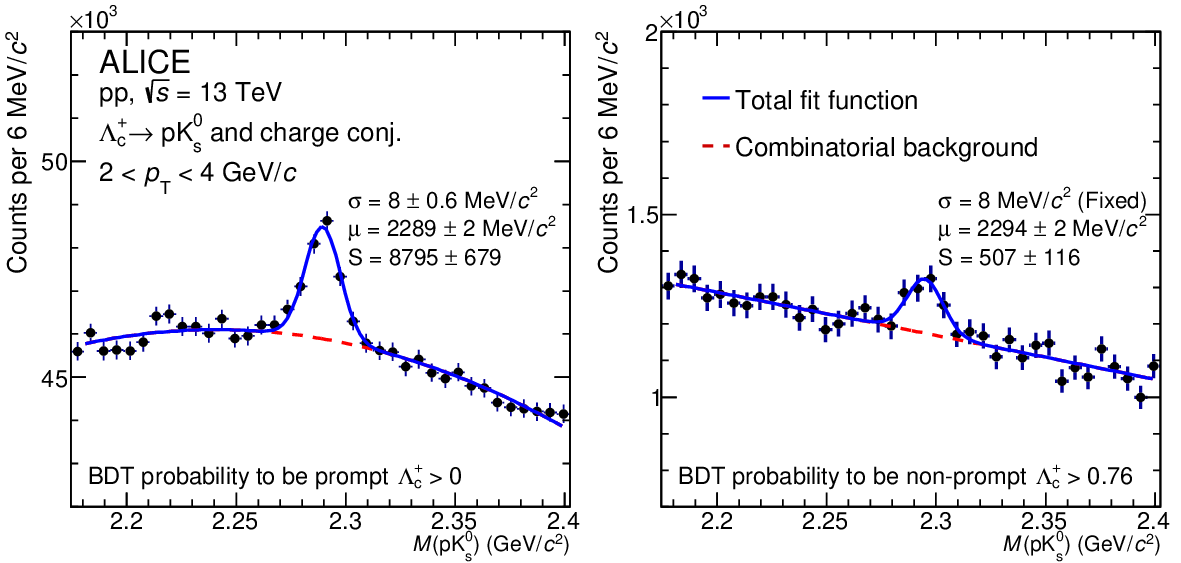}
\includegraphics[width=0.8\textwidth]{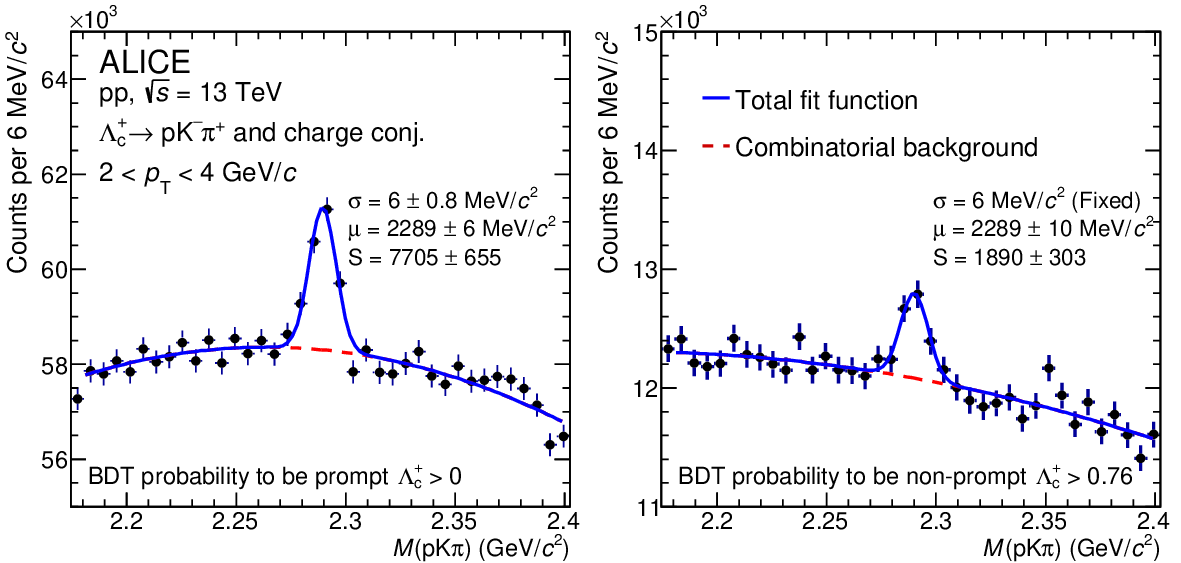}
\caption{Invariant-mass distributions of the \Dzero- and \Lc-hadron candidates and their charge conjugates produced in pp collisions at $\sqrt{s}=13~\TeV$ and reconstructed at midrapidity, shown in a selected \pt range. The values for the Gaussian mean $\mu$, width $\sigma$, and raw yield $S$ are reported. Top row: \DzerotoKpi meson candidates measured in the $2 < \pt < 4~\GeV/c$ interval. Middle row: \LctopKzeros baryon candidates measured in the $2 < \pt < 4~\GeV/c$ interval. Bottom row: \LctopKpi baryon candidates measured in the $2 < \pt < 4~\GeV/c$ interval. The corresponding BDT probability minimum threshold for the candidate selection is reported. The left (right) column corresponds to the prompt (non-prompt) \Dzero- and \Lc- hadron candidates enriched sample.}
\label{fig:mass_D0_Lc}
\end{center}
\end{figure}

The \Dzero-meson and \Lc-baryon raw yields were extracted via binned maximum-likelihood fits to the candidate invariant-mass distributions. The fitting function was composed of a Gaussian term to model the signal and an exponential or polynomial function to model the background. In the case of \Dzero mesons, an additional term was added for the contribution of signal candidates with the wrong $\mathrm{K\mbox{--}\uppi}$ mass assignment (reflections), parameterized from simulated candidates~\cite{ALICE:2019nxm}. To improve the stability of the fits, the widths of the signal peaks were fixed to the values extracted from the fits of the invariant-mass distributions in the prompt enhanced sample, given the naturally larger abundance of prompt compared to non-prompt candidates. Examples of invariant-mass fits with different contributions of signal from beauty-hadron decays in the $2 < \pt < 4~\GeV/c$ interval are shown in Fig.~\ref{fig:mass_D0_Lc} for $\rm D^0$ (top row), \LctopKzeros (middle row), and \LctopKpi (bottom row). The blue solid curves show the total fit function, the red dashed curves show the combinatorial-background contribution, and the green solid lines represent the reflection contribution only for $\rm D^0$. 
 The fits to the invariant-mass distributions of non-prompt (prompt) enhanced samples are shown in each right (left) panel, indicating the corresponding selection applied on the BDT output score related to the probability to be a non-prompt (prompt) charm hadron.

\subsection{Yield corrections and non-prompt fraction estimate}
The non-prompt \Dzero and \Lc $p_{\rm T}$-differential cross sections were obtained in the rapidity interval $|y|<0.5$ as            

\begin{equation}
  \label{eq:CrossSection}
  \frac{\rm d{\sigma}^{\Lc(\Dzero)}}{{\rm d} \pt {\rm d} y}=\frac{1}{\rm BR}\times\frac{1}{2c_{\Delta y}(\pt)\Delta \pt}\times\frac{\fnonprompt \times \Nraw}{\AccEffNP}\times\frac{1}{\Lint},
\end{equation}

where \Nraw is the raw yield (sum of particles and antiparticles),  $c_{\Delta y}(\pt)$ and $\Delta \pt$ represent the width of the rapidity and transverse momentum intervals respectively, BR is the branching ratio of the considered decay mode, the factor of 2 is introduced to obtain the average of particle and antiparticle yields, \fnonprompt is the fraction of non-prompt hadrons in the raw yield, \AccEffNP is the product of the geometrical acceptance and the reconstruction and selection efficiency for non-prompt hadrons, which increases with \pt from 5\% to 25\% depending on the BDT selections and decay channel for \Lc (5\% to 40\% for \Dzero), and \Lint is the integrated luminosity. The correction factors (\AccEff) for the detector acceptance and the signal reconstruction and selection efficiency were determined using the aforementioned MC simulations.

A data-driven procedure based on the construction of data samples with different abundances of prompt and non-prompt candidates was used to estimate the fraction \fnonprompt of non-prompt \Dzero and non-prompt \Lc hadrons. A set of raw yields (\rawY{i}) can be obtained by varying the selection on the BDT output, which is related to the candidate's probability to be a non-prompt \Dzero meson or a non-prompt \Lc baryon. These raw yields are sensitive to the corresponding (\AccEff) of prompt and non-prompt \Dzero or \Lc hadrons as follows

\begin{equation}
   \effP{i}\times\Np +  \effNP{\rm i}\times\Nnp - \rawY{i} = \delta_\mathrm{i},
\label{eq:eq_set}
\end{equation}

where $\delta_\mathrm{i}$ represents a residual that accounts for the equation not holding exactly due to the uncertainties of \rawY{i}, \effNP{\rm i}, and \effP{i}. In the case of $n\geq 2$ sets, a $\chi^2$ function can be defined based on Eq.~\ref{eq:eq_set}, which can be minimized to obtain the corrected yields of prompt (\Np) and non-prompt (\Nnp) \Lc (\Dzero) hadrons as explained in Ref.~\cite{ALICE:2021mgk}. One of the $n$ sets with a high non-prompt component (larger than 50\%) was selected as a working point (default), and the corresponding $f_\mathrm{non\text{-}prompt\text{,default}}$ fraction (more details in Ref.~\cite{ALICE:2021mgk}) can be calculated as 

\begin{equation}
\label{eq:fnpromptSystem}
  f_\mathrm{non\text{-}prompt\text{,default}} = \frac{\effNP{\rm default} \times \Nnp}{\effNP{\rm default} \times \Nnp+\effP{\rm default} \times \Np}.
\end{equation}

\begin{figure}[!tb]
\begin{center}
\includegraphics[width=1.\textwidth]{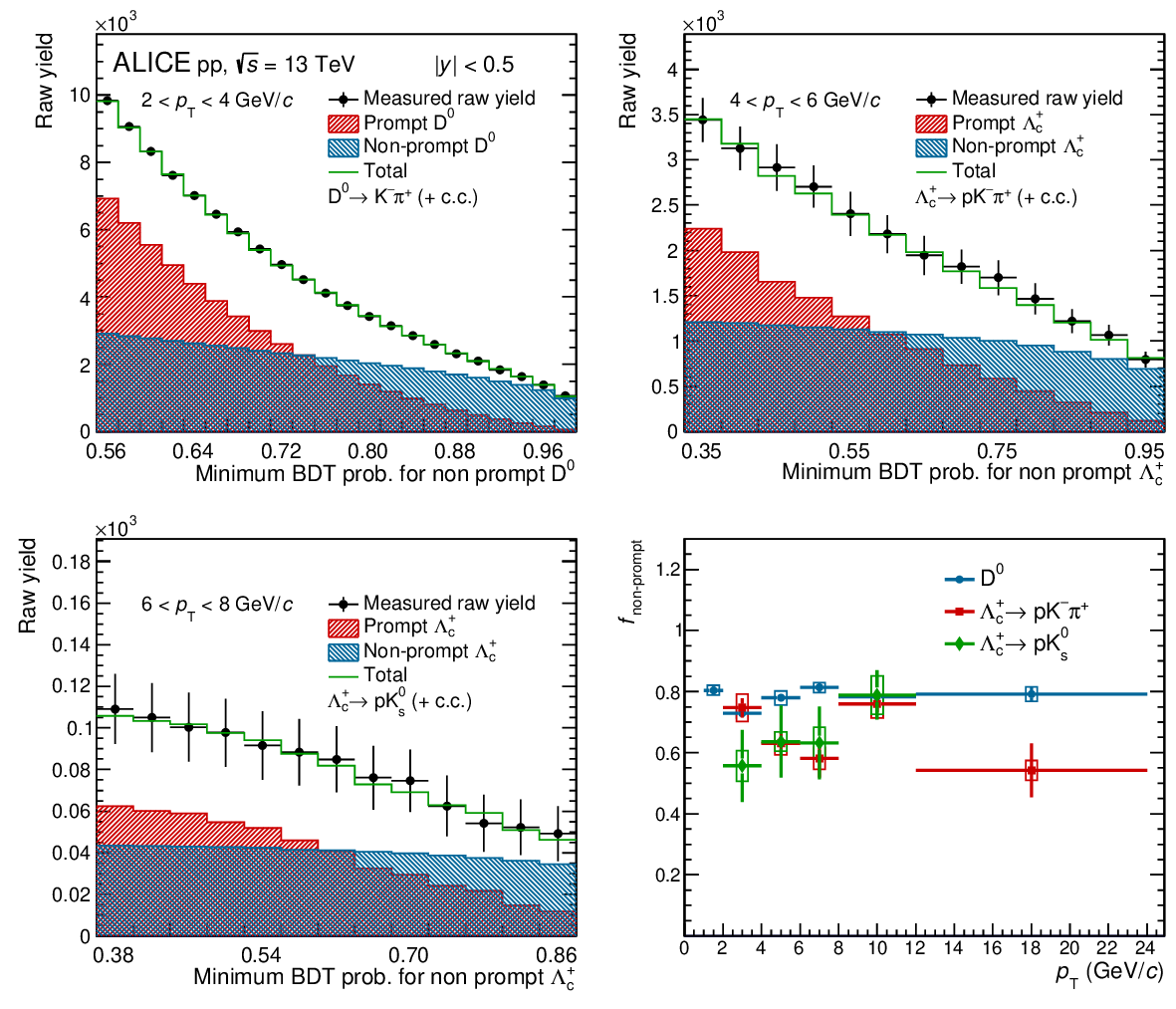}
\caption{Raw-yield distribution as a function of the BDT-based selection employed in the $\chi^{2}$-minimization procedure adopted for the determination of \fnonprompt of \Dzero in $2 < \pt < 4 ~\GeV/c$ (top left panel), \LctopKpi in $4 < \pt < 6 ~\GeV/c$ (top right panel), and \LctopKzeros in $6 < \pt < 8 ~\GeV/c$ (bottom left panel). Bottom right panel: \fnonprompt fraction as function as \pt obtained for the set of selection criteria adopted in the analysis of for non-prompt \Dzero, \LctopKpi, and \LctopKzeros hadrons.}
\label{fig:cut_var_D0_Lc}
\end{center}
\end{figure}

Figure~\ref{fig:cut_var_D0_Lc} shows an example of raw-yield distribution as a function of the BDT-based selection employed in the minimization procedure for \Dzero in $2 < \pt < 4~\GeV/c$ (top left panel), \LctopKpi in $4 < \pt < 6~\GeV/c$ (top right panel), and \LctopKzeros in $6 < \pt < 8~\GeV/c$ (bottom left panel). The black markers are the measured raw yields corresponding to a selection on the BDT output related to the candidate's probability of being a non-prompt \Dzero (\Lc) meson (baryon).
The leftmost data point of each distribution corresponds to the loosest applied selection, while the
rightmost one corresponds to the tightest selection, which preferentially selects non-prompt
candidates. The prompt and non-prompt components, obtained for each BDT-based selection from the
minimization procedure as $\effP{i} \times \Np$ and $\effNP{\rm i} \times \Nnp$, are represented by the red and blue filled histograms, respectively, while their sum is reported by the green histogram. The \fnonprompt fractions obtained for \Dzero, \LctopKpi, and \LctopKzeros, computed for the default selections with the formula in Eq.~\ref{eq:fnpromptSystem} are reported as a function of \pt in the bottom right panel of Fig.~\ref{fig:cut_var_D0_Lc}.

\section{Systematic uncertainties}
The systematic uncertainties of the non-prompt \Lc and \Dzero cross sections were studied for the different decay channels, which depend on \pt. 
The contributions from the raw-yield extraction were evaluated by repeating the invariant-mass fits, varying the fit interval, the functional form of the background fit function, and the width of the Gaussian function used to model the signal peaks. The latter was varied within the uncertainties obtained from the fits of the invariant-mass distributions of the prompt enhanced sample.
The relative uncertainty from this contribution varies in the range 1--3\% for \Dzero and 4--11\% for \Lc. The uncertainties of the track reconstruction efficiency were estimated by considering the uncertainty due to track quality selections and the uncertainty due to the TPC--ITS track matching efficiency as discussed in Ref.~\cite{ALICE:2021mgk}. It ranges from 3.5\% to 5\% for \Dzero and from 4\% to 7\% for \Lc. The systematic uncertainties of the non-prompt fractions were evaluated by varying the configuration and the number of BDT selections employed in the data-driven method and amounts to 2--3\% for \Dzero and 5--9\% for \Lc. The selection efficiency uncertainties, ranging from 2\% to 4\% for \Dzero and 4\% to 10\% for \Lc, were studied by repeating the analyses using different BDT working points. The systematic uncertainties of the PID selection efficiency were found to be negligible, similar to what observed for prompt charm hadrons~\cite{ALICE:2021npz}. The systematic effects due to a possible difference between the real and simulated charm and beauty hadron \pt spectra, were estimated by evaluating the selection efficiency after reweighting the \pt shape from the \pythiaprec event generator to match the one from \fonll calculations~\cite{Cacciari:1998it,Cacciari:2001td,Cacciari:2012ny} for prompt and non-prompt \Dzero. For the \Lc
the reweighting was defined to match the \pt shape of \Dzero and B mesons from \fonll multiplied by the $\Lc/\Dzero$ yield ratio from ALICE~\cite{ALICE:2021rzj} and the $\Lb/\mathrm {B}$ yield ratio from LHCb~\cite{LHCb:2019fns}, respectively. The weights were applied to the \pt distributions for prompt hadrons and to the mother beauty-hadron particles in the case of non-prompt hadrons. The systematic uncertainty from this contribution varies in the range \mbox{1--4\%} for \Dzero and \mbox{4--5\%} for \Lc. In addition, the imperfect apparatus material budget description in the MC simulation, particularly relevant for the effects of the absorption of protons, might result in a bias in the estimation of the \Lc efficiencies. It was evaluated by comparing the corrected yields of charged pions, kaons, and protons using a standard MC production and one with the material budget increased artificially by 10\%. The assigned systematic uncertainty is  2\%. Further \pt-independent uncertainties from the BR~\cite{ParticleDataGroup:2022pth} and the luminosity~\cite{ALICE:2021leo} were considered. 
The total uncertainties, 5--8\% for \Dzero and 12--17\% for \Lc, were calculated as the quadratic sum of the contributions of the different sources.

%% file: results.tex
\section{Results}
\subsection{Production cross sections}

\begin{figure*}[!t]
\begin{center}
\includegraphics[width=0.49\textwidth]{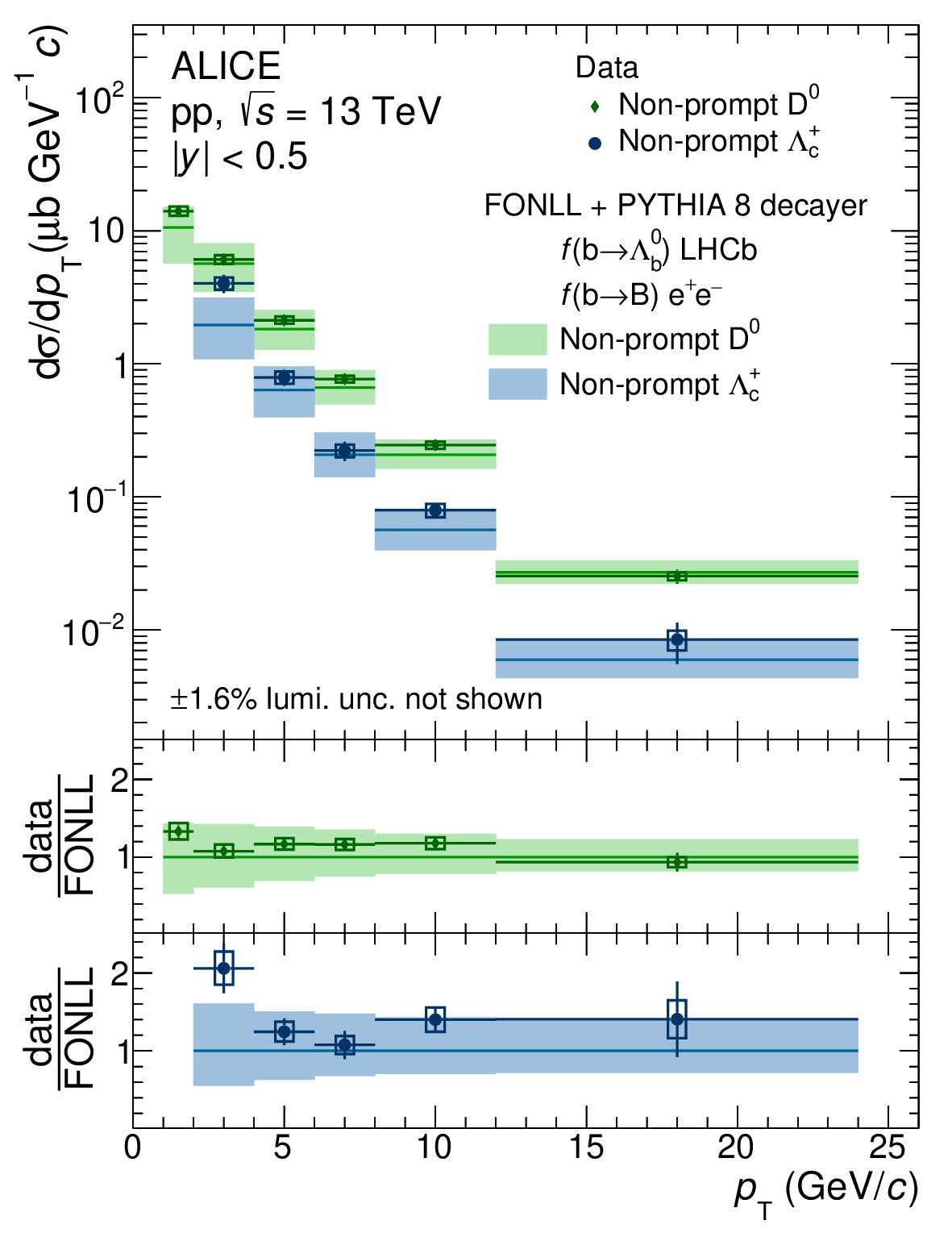}
\includegraphics[width=0.49\textwidth]{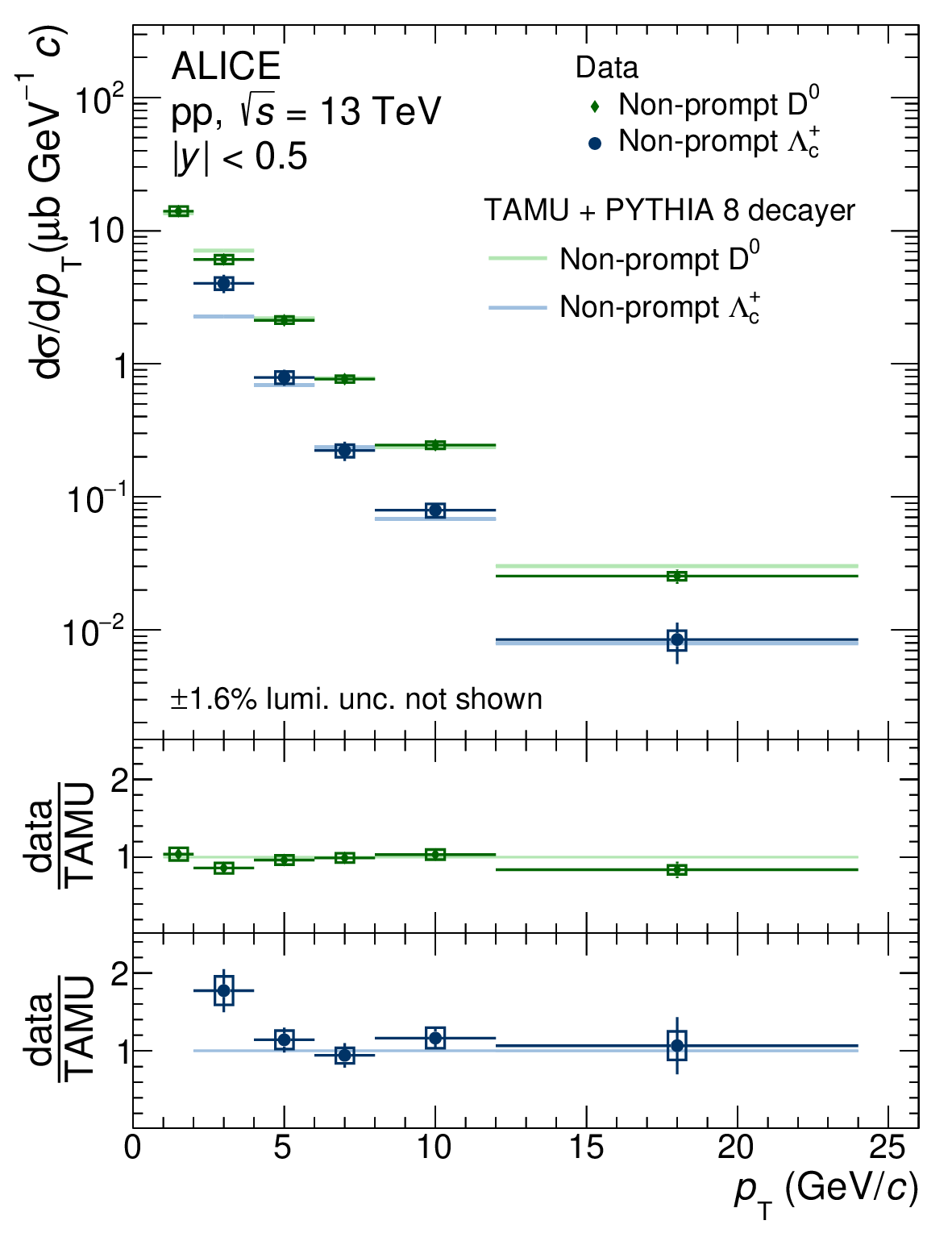}
\caption{\pt-differential production cross sections of non-prompt \Dzero and \Lc hadrons in pp collisions at ${\s=13~\TeV}$ compared with predictions obtained with \fonll calculations~\cite{Cacciari:1998it,Cacciari:2001td,Cacciari:2012ny} adopting \fbtoB and \fbtoLb fragmentation fractions measured in \ee collisions~\cite{ParticleDataGroup:2022pth} and LHCb Collaboration~\cite{LHCb:2019fns} (left panel) and the \tamu model~\cite{He:2022tod} (right panel) combined with \pythia~\cite{Sjostrand:2014zea,Skands:2014pea} for the \HbtoHc decay kinematics.} 
\label{fig:xsec} 
\end{center}
\end{figure*}

The \pt-differential production cross sections of non-prompt \Lc baryons and \Dzero mesons are shown in Fig.~\ref{fig:xsec}.
The non-prompt \Lc cross section was obtained by computing a weighted average of the results from the analyses of the \LctopKzeros and \LctopKpi decay channels, using the inverse of the quadratic sum of the relative statistical and uncorrelated systematic uncertainties as weights. The systematic uncertainties related to the tracking, luminosity, and generated \pt spectrum in the MC simulations are treated as correlated between the two decay channels, the uncertainty of the branching ratios as partially correlated as described in Ref.~\cite{Workman:2022ynf}, while all the other sources of systematic uncertainties are considered uncorrelated.
The data points are compared with theoretical models based on the B-meson cross section predicted by \fonll calculations in the left panel and to the \tamu statistical hadronization model~\cite{He:2022tod} in the right panel. 
In the \fonll-based predictions, the beauty-quark fragmentation fraction to B mesons, \fbtoB, was taken from \ee collisions~\cite{Workman:2022ynf}. For the \Lb baryon, $\fbtoBee$ $\times$ $(\fbtoLb/\fbtoB)_{\mathrm{LHCb}}$ was used, where the \pt-differential ratio \Lb$/$B was measured by the LHCb Collaboration in pp collisions~\cite{LHCb:2019fns}. It has to be noted that the usage of the \pt-differential fragmentation fractions ratio $(\fbtoLb/\fbtoB)_{\mathrm{LHCb}}$ combined with the \fbtoB from \ee collisions enlarges the total beauty production cross section compared to the one predicted by \fonll calculations.
In the \tamu model instead, the branching fractions of beauty quarks to the different hadron species are assumed to follow the relative thermal densities calculated with the statistical hadronization model and an enriched set of heavy-flavor hadron states is obtained from the relativistic-quark model~\cite{Ebert:2011kk}. The \pt distribution is obtained from the one of beauty quarks, convoluted to the fragmentation functions as implemented in \fonll calculations. In this case, the \bbbar production cross section at midrapidity is a parameter of the model, fixed to $\de\sigma_\bbbar/\de y|_{y=0} = 85.3~\mub$.

The resulting beauty-hadron cross sections of both models were then folded with the \HbtoHc decay kinematics (where \Hc and \Hb denote a generic hadron species containing either a charm or beauty quark) and branching ratios provided by the \pythia decayer, in order to obtain the non-prompt \Dzero and \Lc cross sections. In the left panel, the uncertainties in the model are those of the \fonll calculation, which arise from the choice of the normalization and factorization scales, and the mass of the beauty quark, combined with the uncertainties of the \cteq PDFs. The non-prompt \Dzero cross section is in agreement with \fonll $+$ \pythia and \tamu $+$ \pythia predictions over the whole \pt range, while the non-prompt \Lc cross section shows a hint of underestimation at low \pt ($2<\pt<4~\GeV/c$) by both models.

\begin{table}[!b]
\caption{Production cross sections at midrapidity per unit of rapidity $(\de\sigma/\de y)_\mathrm{|y|<0.5}$ in pp collisions at $\s=13~\TeV$ of non-prompt \Lc and \Dzero hadrons and \bbbar pairs $(\de\sigma_{\bbbar}/\de y)_\mathrm{|y|<0.5}$ compared to pQCD calculations~\cite{Cacciari:1998it,Cacciari:2001td,Cacciari:2012ny,Catani:2020kkl}.}
\centering
\renewcommand*{\arraystretch}{1.2}
\begin{tabular}[t]{l|>{\centering}p{0.4\linewidth}|>{\centering}p{0.21\linewidth}|>{\centering\arraybackslash}p{0.21\linewidth}}
\toprule
& Measurement (\mub) & \fonll (\mub)~\cite{Cacciari:1998it,Cacciari:2001td,Cacciari:2012ny} & \nnlo (\mub)~\cite{Catani:2020kkl}\\
\midrule
\Dzero & $41.3\pm1.5(\mathrm{stat})\pm2.9(\mathrm{syst})^{+0.3}_{-1.6} (\mathrm{extrap})$ & $34\pm14$ & -\\
\Lc & $19.1\pm2.3(\mathrm{stat})\pm1.7(\mathrm{syst})^{+1.0}_{-1.5} (\mathrm{extrap})$ & $11_{-5}^{+7}$ & -\\
\bbbar & $83.1 \pm 3.5 (\mathrm{stat}) \pm 5.4(\mathrm{syst}) ^{+12.3}_{-3.2} (\mathrm{extrap})$ & $65_{-26}^{+28}$ & $72^{+22}_{-20}$\\
\bottomrule
\end{tabular}
\label{tab:xsec}	
\end{table}

The measured visible cross sections of non-prompt \Lc and \Dzero hadrons were computed by integrating the measured \pt-differential cross sections in the measured \pt range. All the systematic uncertainties were propagated as fully correlated among the measured \pt intervals, except for the raw-yield extraction uncertainty. As the \pt-differential cross sections predicted by \fonll $+$ \pythia were found to be compatible with the measurements, they were assumed to provide an accurate description of the \pt shape also outside of the measured \pt range. Therefore, the visible cross section was then extrapolated to the full \pt range, using an extrapolation factor computed as the ratio of the \pt-integrated cross sections predicted by \fonll $+$ \pythia integrated over $\pt > 0$ and that in the measured \pt interval. The systematic uncertainty of the extrapolation factors was computed considering (i) the \fonll uncertainties, (ii) the \fbtoHb fragmentation fractions uncertainties, and (iii) the branching ratios uncertainties of the \HbtoHc decays. The second source was estimated by using different sets of beauty fragmentation fractions (from \ee, \ppbar collisions~\cite{Workman:2022ynf}, or those measured by LHCb~\cite{LHCb:2019fns}), while for the third one the branching ratios implemented in \pythia were reweighted in order to reproduce the measured values reported in Ref~\cite{Workman:2022ynf}. The resulting extrapolation factors are $\alpha_\mathrm{extrap}^{\Dzero}=1.241^{+0.009}_{-0.047}$ and $\alpha_\mathrm{extrap}^{\Lc}=1.847^{+0.108}_{-0.152}$ for \Dzero mesons and \Lc baryons, respectively. The \pt-integrated cross sections are reported in Table~\ref{tab:xsec} and compared to \fonll $+$ \pythia calculations, which describe the measurements within the uncertainties.

Similarly, the \bbbar production cross section per unit of rapidity at midrapidity was obtained summing the visible cross sections previously computed and then using an extrapolation factor to account for the unmeasured \pt regions and hadrons. This factor was computed as the ratio of the beauty cross section and the visible cross section of a non-prompt charm hadron, estimated with \fonll $+$ \pythia as follows
\begin{equation}
    \alpha_\mathrm{extrap}^{\bbbar} = \frac{\de\sigma_{\bbbar}^\fonll/\de y|_{|y|<0.5}}{\de\sigma_{\Hc \leftarrow\mathrm{b}}^{\fonll+\pythia}/\de y|_{|y|<0.5} (\pt^\mathrm{min~\Hc}<\pt<\pt^\mathrm{max~\Hc})}.
\end{equation}
The extrapolation factor for the \Dzero meson was found to be $\alpha_\mathrm{extrap}^{\bbbar,~\Dzero} = 2.106^{+0.366}_{-0.014}$, while the one for \Lc baryons $\alpha_\mathrm{extrap}^{\bbbar,~\Lc} = 10.98^{+0.87}_{-1.34}$. The systematic uncertainty of the extrapolation factor includes the same sources considered for the extrapolation of the single-hadron production cross sections. In addition, a correction due to the difference between the rapidity distributions of beauty quarks and beauty hadrons, and between the \bbbar pairs and beauty quarks was applied. The first factor was evaluated to be unity in the relevant rapidity range based on \fonll calculations with 1\% uncertainties evaluated from the difference between \fonll and \pythia. The second correction factor is the ratio $(\de\sigma_{\bbbar}/\de y)/(\de\sigma_\mathrm{b}/\de y) = 1.06\pm0.01$ in $|y|<$~0.5, which was estimated from \powheg simulations~\cite{Alioli:2010xd}. The uncertainty was assigned by varying the factorization and renormalization scales in the \powheg calculation and using the \ctnloten~\cite{Lai:2010vv} and \ctnlofourteen~\cite{Dulat:2015mca} PDFs, alternatively to the default one (\cteq). 
The $\de\sigma_\bbbar/\de y$ was computed separately from the measurements of non-prompt \Dzero and \Lc hadrons were then averaged using the inverse of the quadratic sum of the absolute statistical and uncorrelated systematic uncertainties as weights. The systematic uncertainties related to the tracking uncertainty and the extrapolation uncertainties related to FONLL and the beauty fragmentation fractions were treated as fully correlated among the two hadron species, while all the other sources as uncorrelated. The resulting \bbbar production cross section per unit of rapidity at midrapidity is compatible with the predictions from \fonll and \nnlo calculations, as reported in Table~\ref{tab:xsec}. The \nnlo predictions are however closer to the measurement and have smaller uncertainties than the \fonll ones, as expected by the higher perturbative accuracy. The measurement is also compatible with previous estimates based on the measurements of dielectrons~\cite{ALICE:2018gev, ALICE:2020mfy} and non-prompt \Jpsi mesons~\cite{ALICE:2021edd}.

\begin{figure}[!tb]
\begin{center}
\includegraphics[width=0.8\textwidth]{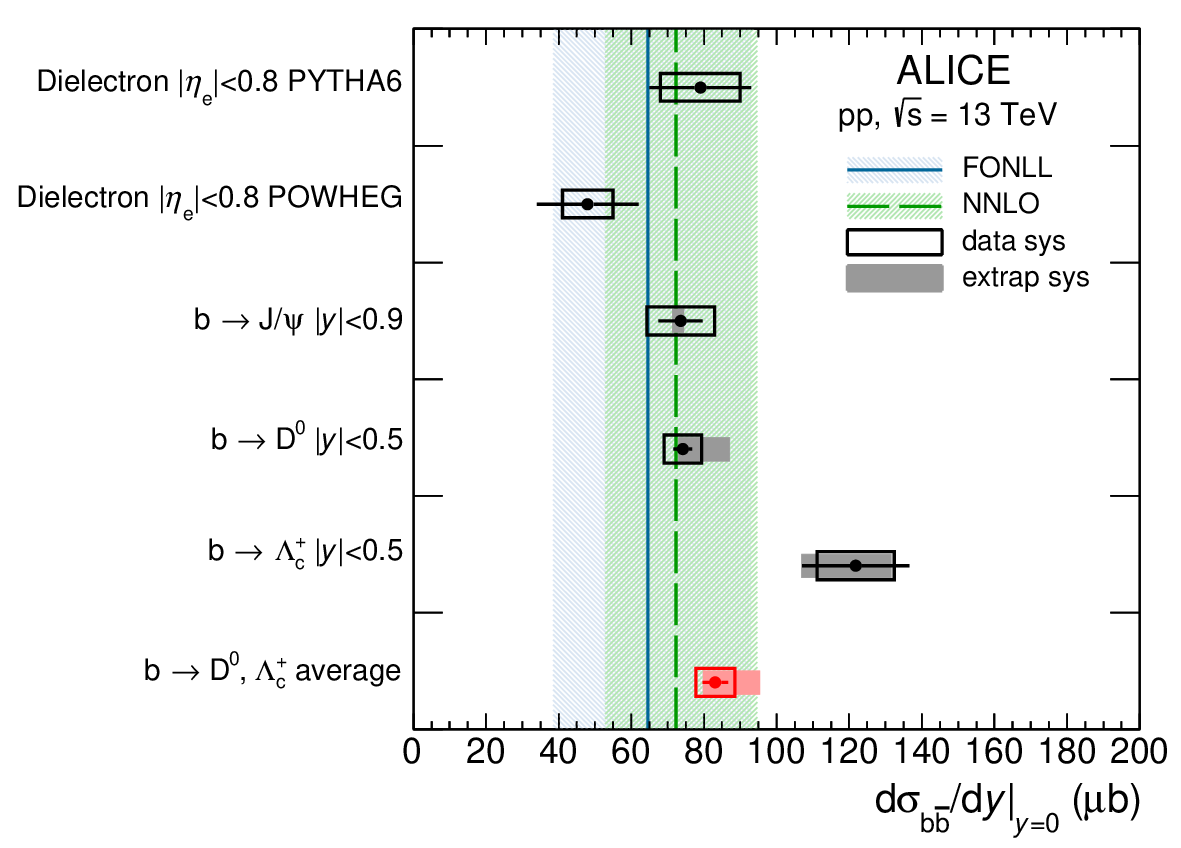}
\caption{Beauty production cross section per rapidity unit at midrapidity obtained from from dielectron~\cite{ALICE:2020mfy} and non-prompt \Jpsi~\cite{ALICE:2021edd}, \Dzero, and \Lc hadrons measured in pp collisions at $\s=13~\TeV$ compared to FONLL~\cite{Cacciari:1998it,Cacciari:2001td,Cacciari:2012ny} and NNLO~\cite{Catani:2020kkl} predictions. The average $\de\sigma_\bbbar/\de y$ of the estimates from the \Dzero and \Lc hadrons is also reported.}
\label{fig:bbbar}
\end{center}
\end{figure}

Figure~\ref{fig:bbbar} shows the \bbbar production cross section per unit rapidity at midrapidity estimated from the production cross sections of non-prompt \Dzero and \Lc hadrons in pp collisions at $\s=13~\TeV$ compared to the previous values based on dielectron and non-prompt \Jpsi-meson measurements. The experimental results are also compared with the predictions provided by \fonll and \nnlo perturbative QCD calculations. 

\subsection{Baryon-to-meson ratios}
\begin{figure*}[!t]
\begin{center}
\includegraphics[width=0.49\textwidth]{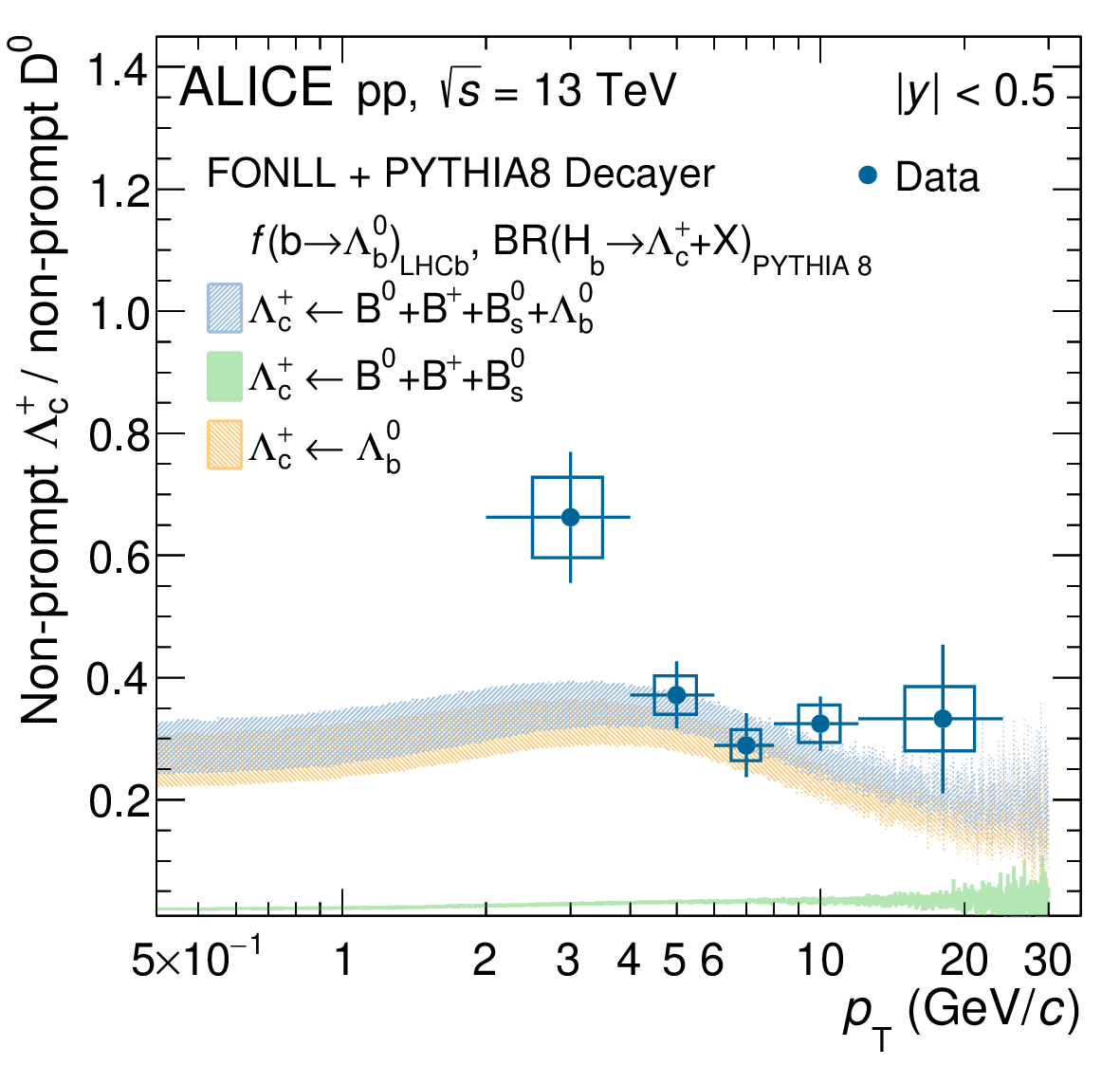}
\includegraphics[width=0.49\textwidth]{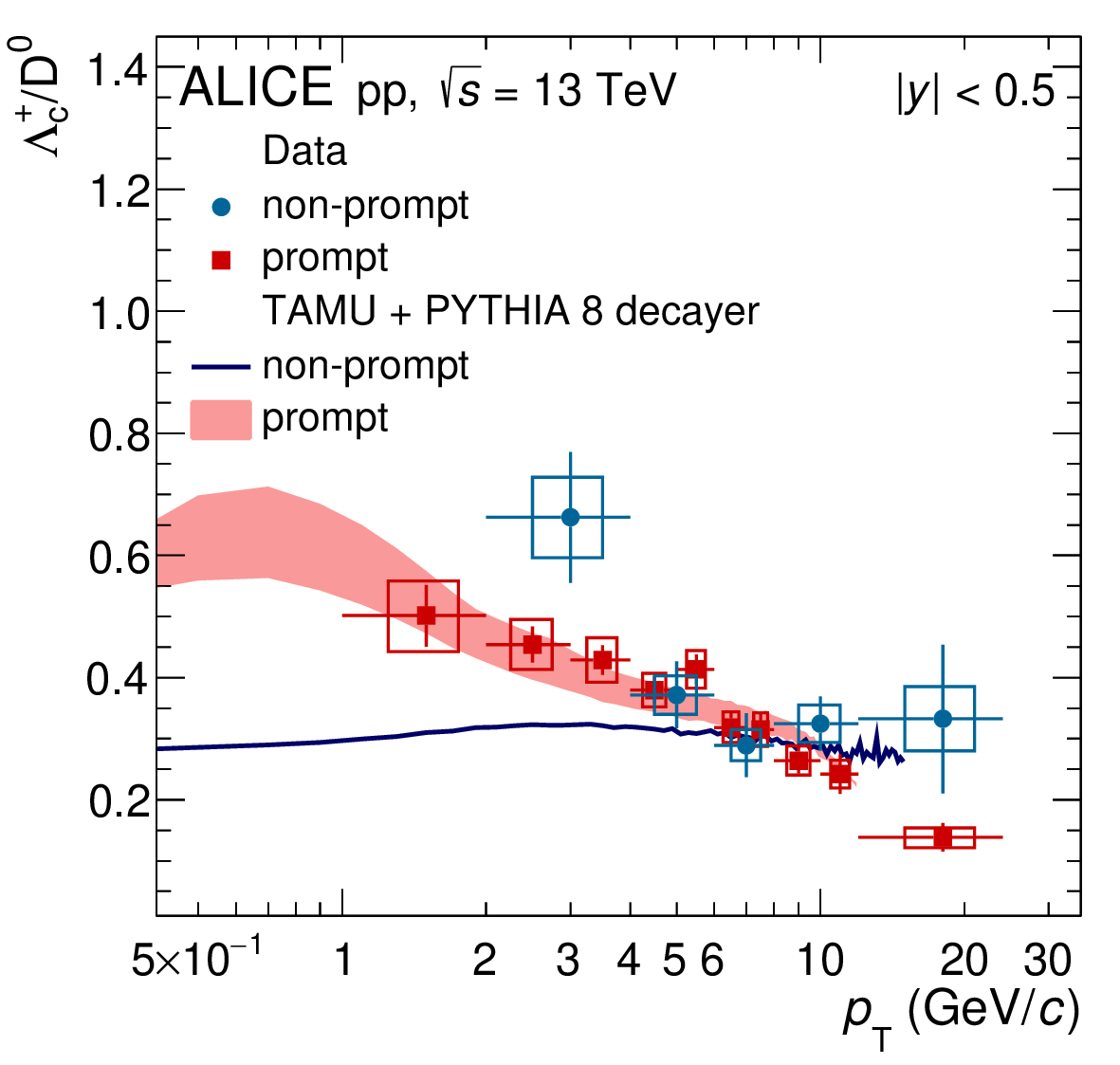}
\caption{Left: \pt-differential ratios of non-prompt \Lc- and \Dzero-hadron cross sections compared with predictions obtained with \fonll calculations~\cite{Cacciari:1998it,Cacciari:2001td,Cacciari:2012ny} and \pythia~\cite{Sjostrand:2014zea,Skands:2014pea} for the \HbtoHc decay kinematics. The contributions from beauty mesons and from the \Lb baryon are depicted separately. Right: \pt-differential ratios of prompt~\cite{ALICE:2021rzj} and non-prompt \Lc- and \Dzero-hadron cross sections compared with predictions obtained with the \tamu model~\cite{He:2022tod,He:2019tik} and \pythia for the \HbtoHc decay kinematics.} 
\label{fig:ratio_fonll_tamu} 
\end{center}
\end{figure*}

The ratio of the \pt-differential production cross sections of non-prompt \Lc and \Dzero hadrons is shown in Fig.~\ref{fig:ratio_fonll_tamu}. In the left panel, the data are compared with theoretical predictions obtained with \fonll calculations~\cite{Cacciari:1998it,Cacciari:2001td,Cacciari:2012ny} and \pythia~\cite{Sjostrand:2014zea,Skands:2014pea} for the description of the decay kinematics and branching ratios. They are obtained using fragmentation fractions from \ee collisions~\cite{ParticleDataGroup:2022pth} for the B mesons and the $\fbtoLb/\fbtoB$ fragmentation fraction ratio measured by the LHCb Collaboration~\cite{LHCb:2019fns}. The contributions of non-prompt \Lc baryons originating from B mesons and \Lb baryons are reported separately to show that the largest contribution is represented by the beauty baryons, while the B mesons contribute only marginally to the non-prompt \Lc production cross section. Hence, it is possible to inquire the hadronization of beauty quarks into \Lb baryons through the non-prompt \Lc. In the right panel of the same figure the data are compared with the \pt-differential ratio between prompt \Lc and \Dzero hadrons. The two measurements are compatible in their common \pt range, with a tension of less than two standard deviations in $2<\pt<4~\GeV/c$, where the ratio of non-prompt hadrons is higher than the one of promptly produced hadrons. The experimental data are also compared to the predictions obtained with the \tamu model combined with \pythia to describe the \HbtoHc decay kinematics, in the case of non-prompt production. The prediction for prompt charm hadrons has an error band representing the uncertainty on the BR of excited charm baryons decaying into \Lc, not included in the one for non-prompt hadrons. 
The measured non-prompt $\Lc/\Dzero$ ratio is rather well described for $\pt>4~\GeV/c$ given the current uncertainties, while it is underestimated for $2<\pt<4~\GeV/c$. The prompt charm and beauty ratios are described by the \tamu calculations within the uncertainties for the whole measured $\pt$ interval.

\begin{figure*}[!t]
\begin{center}
\includegraphics[width=0.95\textwidth]{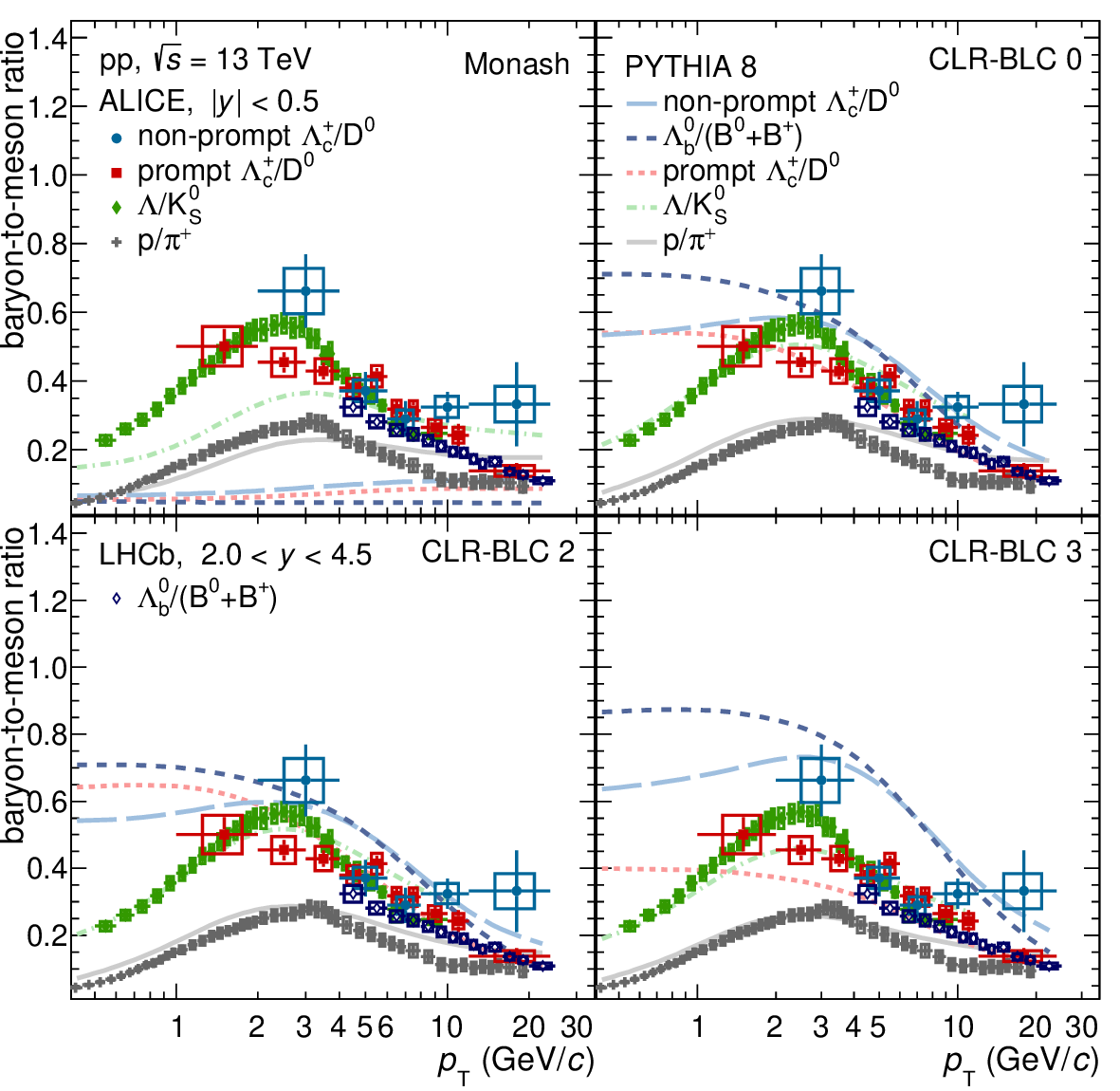}
\caption{Non-prompt $\Lc/\Dzero$, prompt $\Lc/\Dzero$~\cite{ALICE:2021rzj}, $\lmb/\kzero$~\cite{ALICE:2020jsh}, and p$/\pip$~\cite{ALICE:2020jsh} ratios measured in pp collisions at $\s=13~\TeV$ at midrapidity ($|y|<0.5$) compared with the $\Lb/(\Bzero+\Bplus)$ ratio measured by the LHCb Collaboration at forward rapidity ($2.5<y<4$)~\cite{LHCb:2019fns} and with predictions obtained with the \pythia MC generator with the Monash 2013 tune~\cite{Sjostrand:2014zea,Skands:2014pea} and the \clrblc modes 0, 2, and 3~\cite{Christiansen:2015yqa} in the corresponding rapidity range with respect to data.} 
\label{fig:ratio_pythia} 
\end{center}
\end{figure*}

Figure~\ref{fig:ratio_pythia} shows the \pt-differential non-prompt $\Lc/\Dzero$ yield ratio at midrapidity ($|y|<0.5$) in pp collisions at $\s=13~\TeV$ compared with the measurements of prompt $\Lc/\Dzero$~\cite{ALICE:2021rzj}, $\lmb/\kzero$~\cite{ALICE:2020jsh}, and p$/\pip$~\cite{ALICE:2020jsh} ratios at the same energy and rapidity interval, and with the $\Lb/(\Bzero+\Bplus)$ yield ratio measured by LHCb at forward rapidity ($2.5<y<4$). The $\Lb/(\Bzero+\Bplus)$ ratio is a bit lower than the one between non-prompt $\Lc$ and $\Dzero$ mesons, however it has to be considered that the normalization is slightly different. In the LHCb result the production cross sections of \Bzero and \Bplus mesons, i.e. the total yield of non-strange B mesons is used, while the non-prompt $\Dzero$, used in this ratio, accounts for about 70\% of the non-strange D mesons. Also the fraction of \Bzero and \Bplus mesons decaying to \Lc and \Ds, as well as the \Lb and \Bs hadrons decaying to \Dzero mesons influence the ratio. In addition, in the non-prompt $\Lc/\Dzero$ ratio, the $\HbtoHc$ decay kinematics is expected to slightly modify the \pt dependence compared to the one of the ratio between beauty hadrons. Interestingly, all the measurements for beauty, charm, and strange hadrons show a similar trend as a function of \pt and are compatible within the uncertainties. The p$/\pip$ production ratio also features a similar \pt dependence, however it is lower in absolute terms.
The experimental values are compared with the corresponding predictions obtained with \pythia simulations, using different tunes and the same rapidity ranges of the experimental results. In the top-left panel, the results obtained with the Monash 2013 tune~\cite{Skands:2014pea}, which implements a fragmentation process tuned to reproduce the measurements in \ee collisions, is reported. Here, all the baryon-to-meson ratios are underestimated by \pythia~except for the p$/\pip$ ratio for which the model prediction is rather good at low \pt. A better agreement is instead obtained with the \clrblc tunes (i.e.~Mode 0, 2, and 3), shown in the other three panels of Fig.~\ref{fig:ratio_pythia}. These three tunes are characterized by different constraints on the time dilation and causality, as defined in Ref.~\cite{Christiansen:2015yqa}. The time parameters are relevant in this model, because two strings can reconnect if they are able to resolve each other during the time between their formation and hadronization, taking also into account time-dilation effects caused by relative boosts. The Mode 0 and 2 settings, reported in the top-right and bottom-left panels of Fig.~\ref{fig:ratio_pythia} respectively, predict a similar baryon-to-meson ratio for the strange, charm, and beauty flavors for $\pt>2~\GeV/c$ and a significantly higher ratio for heavy-flavor hadrons than strange hadrons for lower \pt (e.g., a factor three is predicted at \pt$\approx$ 400 MeV/$c$). Despite the agreement with the data is significantly improved compared to the Monash tune, the measurements of beauty hadrons are overestimated for $\pt\lesssim10~\GeV/c$. 
Instead, the Mode 3 (bottom-right panel of Fig.~\ref{fig:ratio_pythia}) underestimates the ratio for charm hadrons for $\pt\lesssim12~\GeV/c$ and overestimates that of beauty hadrons in the same \pt interval, quantitatively more than the other two \clrblc modes. The features, observed in all comparisons with \pythia tunes, indicate that more precise measurements of the baryon-to-meson ratios, especially those including beauty-hadron measurements at very low \pt ($\pt<2~\GeV/c$) are crucial for tuning the model parameters involving the reconnection of quarks via junction topologies and to possibly validate this as the mechanism responsible of the baryon enhancement observed in hadron collisions compared to \ee collisions. It is worth pointing out that other theoretical models are proposed to describe the enhancement, based on different hadronization mechanisms (e.g. recombination).

The \pt-integrated non-prompt $\Lc/\Dzero$ ratio was computed by dividing the \pt-integrated cross sections reported in Table~\ref{tab:xsec}. All the systematic uncertainties, except for those related to the tracking efficiency, were propagated as uncorrelated in the ratio. The resulting value is compatible with the one measured for promptly produced particles and significantly higher than that measured in \ee collisions at LEP~\cite{Gladilin:2014tba}. All the values are reported in Table~\ref{tab:ratios}.

\begin{table}[!tb]
\caption{\pt-integrated $\Lc/\Dzero$ production ratio measured at midrapidity ($|y|<0.5$) in pp collisions at $\s=13~\TeV$ and in \ee collisions at LEP~\cite{Gladilin:2014tba} for prompt and non-prompt production.}
\centering
\renewcommand*{\arraystretch}{1.2}
\begin{tabular}[t]{l|>{\centering}p{0.45\linewidth}|>{\centering\arraybackslash}p{0.2\linewidth}}
\toprule
& ALICE & LEP average~\cite{Gladilin:2014tba}\\
\midrule
prompt $\Lc/\Dzero$ & $0.49\pm 0.02(\text{stat})^{+0.05}_{-0.04}(\mathrm{syst})^{+0.01}_{-0.03}(\mathrm{syst})$~\cite{ALICE:2021npz} & $0.105\pm 0.013$\\
non-prompt $\Lc/\Dzero$ & $0.47\pm 0.06 (\text{stat})\pm 0.04 (\text{syst}) ^{+0.03}_{-0.04}(\mathrm{extrap})$ & $0.124\pm 0.016$\\
\bottomrule
\end{tabular}
\label{tab:ratios}	
\end{table}

\section{Conclusions}
In summary, the \pt-differential and \pt-integrated production cross sections of non-prompt \Lc and \Dzero hadrons were measured for the first time at midrapidity in pp collisions at $\s=13~\TeV$. The results are compatible with the theoretical models based on \fonll calculations with the \fbtoLb and \fbtoB fragmentation fractions measured by LHCb and at \ee, respectively, suggesting a similar beauty-baryon enhancement at forward and midrapidity in pp collisions. Furthermore, the results are in agreement with the \tamu statistical hadronization model for the relative abundances of different beauty hadron species. The extrapolated \bbbar production cross section at midrapidity per unit of rapidity is found to be compatible with pQCD calculations with \fonll and \nnlo accuracy. The measured baryon-to-meson ratios of light flavor, strange, charm, and beauty hadrons show a similar \pt trend. In addition, all ratios, except the p$/\pip$, are significantly higher than the values measured in \ee collisions. The \pt-differential baryon-to-meson ratios have been compared to predictions of the \tamu statistical hadronization model and to the \pythia simulations, that include the color-reconnection mechanism in the string fragmentation and indicate that all the flavors have to be considered simultaneously in order to obtain the best tuning of the model parameters involving the reconnection of quarks via junction topologies. This feature asks for more precise results, including a direct measurement of beauty hadrons especially in the same \pt and rapidity range and the $\pt<4~\GeV/c$ region, which can be reached with the data collected in the LHC Run 3 data taking period.

%% file: fa_2023-06-23_Opt_C.tex

The ALICE Collaboration would like to thank all its engineers and technicians for their invaluable contributions to the construction of the experiment and the CERN accelerator teams for the outstanding performance of the LHC complex.
The ALICE Collaboration gratefully acknowledges the resources and support provided by all Grid centres and the Worldwide LHC Computing Grid (WLCG) collaboration.
The ALICE Collaboration acknowledges the following funding agencies for their support in building and running the ALICE detector:
A. I. Alikhanyan National Science Laboratory (Yerevan Physics Institute) Foundation (ANSL), State Committee of Science and World Federation of Scientists (WFS), Armenia;
Austrian Academy of Sciences, Austrian Science Fund (FWF): [M 2467-N36] and Nationalstiftung f\"{u}r Forschung, Technologie und Entwicklung, Austria;
Ministry of Communications and High Technologies, National Nuclear Research Center, Azerbaijan;
Conselho Nacional de Desenvolvimento Cient\'{\i}fico e Tecnol\'{o}gico (CNPq), Financiadora de Estudos e Projetos (Finep), Funda\c{c}\~{a}o de Amparo \`{a} Pesquisa do Estado de S\~{a}o Paulo (FAPESP) and Universidade Federal do Rio Grande do Sul (UFRGS), Brazil;
Bulgarian Ministry of Education and Science, within the National Roadmap for Research Infrastructures 2020-2027 (object CERN), Bulgaria;
Ministry of Education of China (MOEC) , Ministry of Science \& Technology of China (MSTC) and National Natural Science Foundation of China (NSFC), China;
Ministry of Science and Education and Croatian Science Foundation, Croatia;
Centro de Aplicaciones Tecnol\'{o}gicas y Desarrollo Nuclear (CEADEN), Cubaenerg\'{\i}a, Cuba;
Ministry of Education, Youth and Sports of the Czech Republic, Czech Republic;
The Danish Council for Independent Research | Natural Sciences, the VILLUM FONDEN and Danish National Research Foundation (DNRF), Denmark;
Helsinki Institute of Physics (HIP), Finland;
Commissariat \`{a} l'Energie Atomique (CEA) and Institut National de Physique Nucl\'{e}aire et de Physique des Particules (IN2P3) and Centre National de la Recherche Scientifique (CNRS), France;
Bundesministerium f\"{u}r Bildung und Forschung (BMBF) and GSI Helmholtzzentrum f\"{u}r Schwerionenforschung GmbH, Germany;
General Secretariat for Research and Technology, Ministry of Education, Research and Religions, Greece;
National Research, Development and Innovation Office, Hungary;
Department of Atomic Energy Government of India (DAE), Department of Science and Technology, Government of India (DST), University Grants Commission, Government of India (UGC) and Council of Scientific and Industrial Research (CSIR), India;
National Research and Innovation Agency - BRIN, Indonesia;
Istituto Nazionale di Fisica Nucleare (INFN), Italy;
Japanese Ministry of Education, Culture, Sports, Science and Technology (MEXT) and Japan Society for the Promotion of Science (JSPS) KAKENHI, Japan;
Consejo Nacional de Ciencia (CONACYT) y Tecnolog\'{i}a, through Fondo de Cooperaci\'{o}n Internacional en Ciencia y Tecnolog\'{i}a (FONCICYT) and Direcci\'{o}n General de Asuntos del Personal Academico (DGAPA), Mexico;
Nederlandse Organisatie voor Wetenschappelijk Onderzoek (NWO), Netherlands;
The Research Council of Norway, Norway;
Commission on Science and Technology for Sustainable Development in the South (COMSATS), Pakistan;
Pontificia Universidad Cat\'{o}lica del Per\'{u}, Peru;
Ministry of Education and Science, National Science Centre and WUT ID-UB, Poland;
Korea Institute of Science and Technology Information and National Research Foundation of Korea (NRF), Republic of Korea;
Ministry of Education and Scientific Research, Institute of Atomic Physics, Ministry of Research and Innovation and Institute of Atomic Physics and Universitatea Nationala de Stiinta si Tehnologie Politehnica Bucuresti, Romania;
Ministry of Education, Science, Research and Sport of the Slovak Republic, Slovakia;
National Research Foundation of South Africa, South Africa;
Swedish Research Council (VR) and Knut \& Alice Wallenberg Foundation (KAW), Sweden;
European Organization for Nuclear Research, Switzerland;
Suranaree University of Technology (SUT), National Science and Technology Development Agency (NSTDA) and National Science, Research and Innovation Fund (NSRF via PMU-B B05F650021), Thailand;
Turkish Energy, Nuclear and Mineral Research Agency (TENMAK), Turkey;
National Academy of  Sciences of Ukraine, Ukraine;
Science and Technology Facilities Council (STFC), United Kingdom;
National Science Foundation of the United States of America (NSF) and United States Department of Energy, Office of Nuclear Physics (DOE NP), United States of America.
In addition, individual groups or members have received support from:
European Research Council, Strong 2020 - Horizon 2020 (grant nos. 950692, 824093), European Union;
Academy of Finland (Center of Excellence in Quark Matter) (grant nos. 346327, 346328), Finland.

%% file: 2023-06-23-Alice_Authorlist_2023-06-23_Opt_C.tex
\begin{flushleft} 
\small

S.~Acharya\,\orcidlink{0000-0002-9213-5329}\,$^{\rm 128}$, 
D.~Adamov\'{a}\,\orcidlink{0000-0002-0504-7428}\,$^{\rm 87}$, 
G.~Aglieri Rinella\,\orcidlink{0000-0002-9611-3696}\,$^{\rm 33}$, 
M.~Agnello\,\orcidlink{0000-0002-0760-5075}\,$^{\rm 30}$, 
N.~Agrawal\,\orcidlink{0000-0003-0348-9836}\,$^{\rm 52}$, 
Z.~Ahammed\,\orcidlink{0000-0001-5241-7412}\,$^{\rm 136}$, 
S.~Ahmad\,\orcidlink{0000-0003-0497-5705}\,$^{\rm 16}$, 
S.U.~Ahn\,\orcidlink{0000-0001-8847-489X}\,$^{\rm 72}$, 
I.~Ahuja\,\orcidlink{0000-0002-4417-1392}\,$^{\rm 38}$, 
A.~Akindinov\,\orcidlink{0000-0002-7388-3022}\,$^{\rm 142}$
% ...

\section*{Affiliation Notes}

$^{\rm I}$ Deceased\\
$^{\rm II}$ Also at: Max-Planck-Institut fur Physik, Munich, Germany\\
$^{\rm III}$ Also at: Italian National Agency for New Technologies, Energy and Sustainable Economic Development (ENEA), Bologna, Italy\\
$^{\rm IV}$ Also at: Dipartimento DET del Politecnico di Torino, Turin, Italy\\
$^{\rm V}$ Also at: Department of Applied Physics, Aligarh Muslim University, Aligarh, India\\
$^{\rm VI}$ Also at: Institute of Theoretical Physics, University of Wroclaw, Poland\\
$^{\rm VII}$ Also at: An institution covered by a cooperation agreement with CERN\\

\section*{Collaboration Institutes}

$^{1}$ A.I. Alikhanyan National Science Laboratory (Yerevan Physics Institute) Foundation, Yerevan, Armenia\\
$^{2}$ AGH University of Krakow, Cracow, Poland\\
$^{3}$ Bogolyubov Institute for Theoretical Physics, National Academy of Sciences of Ukraine, Kiev, Ukraine\\
$^{4}$ Bose Institute, Department of Physics  and Centre for Astroparticle Physics and Space Science (CAPSS), Kolkata, India\\
$^{5}$ California Polytechnic State University, San Luis Obispo, California, United States\\
$^{6}$ Central China Normal University, Wuhan, China\\
$^{7}$ Centro de Aplicaciones Tecnol\'{o}gicas y Desarrollo Nuclear (CEADEN), Havana, Cuba\\
$^{8}$ Centro de Investigaci\'{o}n y de Estudios Avanzados (CINVESTAV), Mexico City and M\'{e}rida, Mexico\\
$^{9}$ Chicago State University, Chicago, Illinois, United States\\
$^{10}$ China Institute of Atomic Energy, Beijing, China\\
$^{11}$ China University of Geosciences, Wuhan, China\\
$^{12}$ Chungbuk National University, Cheongju, Republic of Korea\\
$^{13}$ Comenius University Bratislava, Faculty of Mathematics, Physics and Informatics, Bratislava, Slovak Republic\\
$^{14}$ COMSATS University Islamabad, Islamabad, Pakistan\\
$^{15}$ Creighton University, Omaha, Nebraska, United States\\
$^{16}$ Department of Physics, Aligarh Muslim University, Aligarh, India\\
$^{17}$ Department of Physics, Pusan National University, Pusan, Republic of Korea\\
$^{18}$ Department of Physics, Sejong University, Seoul, Republic of Korea\\
$^{19}$ Department of Physics, University of California, Berkeley, California, United States\\
$^{20}$ Department of Physics, University of Oslo, Oslo, Norway\\
$^{21}$ Department of Physics and Technology, University of Bergen, Bergen, Norway\\
$^{22}$ Dipartimento di Fisica, Universit\`{a} di Pavia, Pavia, Italy\\
$^{23}$ Dipartimento di Fisica dell'Universit\`{a} and Sezione INFN, Cagliari, Italy\\
$^{24}$ Dipartimento di Fisica dell'Universit\`{a} and Sezione INFN, Trieste, Italy\\
$^{25}$ Dipartimento di Fisica dell'Universit\`{a} and Sezione INFN, Turin, Italy\\
$^{26}$ Dipartimento di Fisica e Astronomia dell'Universit\`{a} and Sezione INFN, Bologna, Italy\\
$^{27}$ Dipartimento di Fisica e Astronomia dell'Universit\`{a} and Sezione INFN, Catania, Italy\\
$^{28}$ Dipartimento di Fisica e Astronomia dell'Universit\`{a} and Sezione INFN, Padova, Italy\\
$^{29}$ Dipartimento di Fisica `E.R.~Caianiello' dell'Universit\`{a} and Gruppo Collegato INFN, Salerno, Italy\\
$^{30}$ Dipartimento DISAT del Politecnico and Sezione INFN, Turin, Italy\\
$^{31}$ Dipartimento di Scienze MIFT, Universit\`{a} di Messina, Messina, Italy\\
$^{32}$ Dipartimento Interateneo di Fisica `M.~Merlin' and Sezione INFN, Bari, Italy\\
$^{33}$ European Organization for Nuclear Research (CERN), Geneva, Switzerland\\
$^{34}$ Faculty of Electrical Engineering, Mechanical Engineering and Naval Architecture, University of Split, Split, Croatia\\
$^{35}$ Faculty of Engineering and Science, Western Norway University of Applied Sciences, Bergen, Norway\\
$^{36}$ Faculty of Nuclear Sciences and Physical Engineering, Czech Technical University in Prague, Prague, Czech Republic\\
$^{37}$ Faculty of Physics, Sofia University, Sofia, Bulgaria\\
$^{38}$ Faculty of Science, P.J.~\v{S}af\'{a}rik University, Ko\v{s}ice, Slovak Republic\\
$^{39}$ Frankfurt Institute for Advanced Studies, Johann Wolfgang Goethe-Universit\"{a}t Frankfurt, Frankfurt, Germany\\
$^{40}$ Fudan University, Shanghai, China\\
$^{41}$ Gangneung-Wonju National University, Gangneung, Republic of Korea\\
$^{42}$ Gauhati University, Department of Physics, Guwahati, India\\
$^{43}$ Helmholtz-Institut f\"{u}r Strahlen- und Kernphysik, Rheinische Friedrich-Wilhelms-Universit\"{a}t Bonn, Bonn, Germany\\
$^{44}$ Helsinki Institute of Physics (HIP), Helsinki, Finland\\
$^{45}$ High Energy Physics Group,  Universidad Aut\'{o}noma de Puebla, Puebla, Mexico\\
$^{46}$ Horia Hulubei National Institute of Physics and Nuclear Engineering, Bucharest, Romania\\
$^{47}$ HUN-REN Wigner Research Centre for Physics, Budapest, Hungary\\
$^{48}$ Indian Institute of Technology Bombay (IIT), Mumbai, India\\
$^{49}$ Indian Institute of Technology Indore, Indore, India\\
$^{50}$ INFN, Laboratori Nazionali di Frascati, Frascati, Italy\\
$^{51}$ INFN, Sezione di Bari, Bari, Italy\\
$^{52}$ INFN, Sezione di Bologna, Bologna, Italy\\
$^{53}$ INFN, Sezione di Cagliari, Cagliari, Italy\\
$^{54}$ INFN, Sezione di Catania, Catania, Italy\\
$^{55}$ INFN, Sezione di Padova, Padova, Italy\\
$^{56}$ INFN, Sezione di Pavia, Pavia, Italy\\
$^{57}$ INFN, Sezione di Torino, Turin, Italy\\
$^{58}$ INFN, Sezione di Trieste, Trieste, Italy\\
$^{59}$ Inha University, Incheon, Republic of Korea\\
$^{60}$ Institute for Gravitational and Subatomic Physics (GRASP), Utrecht University/Nikhef, Utrecht, Netherlands\\
$^{61}$ Institute of Experimental Physics, Slovak Academy of Sciences, Ko\v{s}ice, Slovak Republic\\
$^{62}$ Institute of Physics, Homi Bhabha National Institute, Bhubaneswar, India\\
$^{63}$ Institute of Physics of the Czech Academy of Sciences, Prague, Czech Republic\\
$^{64}$ Institute of Space Science (ISS), Bucharest, Romania\\
$^{65}$ Institut f\"{u}r Kernphysik, Johann Wolfgang Goethe-Universit\"{a}t Frankfurt, Frankfurt, Germany\\
$^{66}$ Instituto de Ciencias Nucleares, Universidad Nacional Aut\'{o}noma de M\'{e}xico, Mexico City, Mexico\\
$^{67}$ Instituto de F\'{i}sica, Universidade Federal do Rio Grande do Sul (UFRGS), Porto Alegre, Brazil\\
$^{68}$ Instituto de F\'{\i}sica, Universidad Nacional Aut\'{o}noma de M\'{e}xico, Mexico City, Mexico\\
$^{69}$ iThemba LABS, National Research Foundation, Somerset West, South Africa\\
$^{70}$ Jeonbuk National University, Jeonju, Republic of Korea\\
$^{71}$ Johann-Wolfgang-Goethe Universit\"{a}t Frankfurt Institut f\"{u}r Informatik, Fachbereich Informatik und Mathematik, Frankfurt, Germany\\
$^{72}$ Korea Institute of Science and Technology Information, Daejeon, Republic of Korea\\
$^{73}$ KTO Karatay University, Konya, Turkey\\
$^{74}$ Laboratoire de Physique Subatomique et de Cosmologie, Universit\'{e} Grenoble-Alpes, CNRS-IN2P3, Grenoble, France\\
$^{75}$ Lawrence Berkeley National Laboratory, Berkeley, California, United States\\
$^{76}$ Lund University Department of Physics, Division of Particle Physics, Lund, Sweden\\
$^{77}$ Nagasaki Institute of Applied Science, Nagasaki, Japan\\
$^{78}$ Nara Women{'}s University (NWU), Nara, Japan\\
$^{79}$ National and Kapodistrian University of Athens, School of Science, Department of Physics , Athens, Greece\\
$^{80}$ National Centre for Nuclear Research, Warsaw, Poland\\
$^{81}$ National Institute of Science Education and Research, Homi Bhabha National Institute, Jatni, India\\
$^{82}$ National Nuclear Research Center, Baku, Azerbaijan\\
$^{83}$ National Research and Innovation Agency - BRIN, Jakarta, Indonesia\\
$^{84}$ Niels Bohr Institute, University of Copenhagen, Copenhagen, Denmark\\
$^{85}$ Nikhef, National institute for subatomic physics, Amsterdam, Netherlands\\
$^{86}$ Nuclear Physics Group, STFC Daresbury Laboratory, Daresbury, United Kingdom\\
$^{87}$ Nuclear Physics Institute of the Czech Academy of Sciences, Husinec-\v{R}e\v{z}, Czech Republic\\
$^{88}$ Oak Ridge National Laboratory, Oak Ridge, Tennessee, United States\\
$^{89}$ Ohio State University, Columbus, Ohio, United States\\
$^{90}$ Physics department, Faculty of science, University of Zagreb, Zagreb, Croatia\\
$^{91}$ Physics Department, Panjab University, Chandigarh, India\\
$^{92}$ Physics Department, University of Jammu, Jammu, India\\
$^{93}$ Physics Program and International Institute for Sustainability with Knotted Chiral Meta Matter (SKCM2), Hiroshima University, Hiroshima, Japan\\
$^{94}$ Physikalisches Institut, Eberhard-Karls-Universit\"{a}t T\"{u}bingen, T\"{u}bingen, Germany\\
$^{95}$ Physikalisches Institut, Ruprecht-Karls-Universit\"{a}t Heidelberg, Heidelberg, Germany\\
$^{96}$ Physik Department, Technische Universit\"{a}t M\"{u}nchen, Munich, Germany\\
$^{97}$ Politecnico di Bari and Sezione INFN, Bari, Italy\\
$^{98}$ Research Division and ExtreMe Matter Institute EMMI, GSI Helmholtzzentrum f\"ur Schwerionenforschung GmbH, Darmstadt, Germany\\
$^{99}$ Saga University, Saga, Japan\\
$^{100}$ Saha Institute of Nuclear Physics, Homi Bhabha National Institute, Kolkata, India\\
$^{101}$ School of Physics and Astronomy, University of Birmingham, Birmingham, United Kingdom\\
$^{102}$ Secci\'{o}n F\'{\i}sica, Departamento de Ciencias, Pontificia Universidad Cat\'{o}lica del Per\'{u}, Lima, Peru\\
$^{103}$ Stefan Meyer Institut f\"{u}r Subatomare Physik (SMI), Vienna, Austria\\
$^{104}$ SUBATECH, IMT Atlantique, Nantes Universit\'{e}, CNRS-IN2P3, Nantes, France\\
$^{105}$ Sungkyunkwan University, Suwon City, Republic of Korea\\
$^{106}$ Suranaree University of Technology, Nakhon Ratchasima, Thailand\\
$^{107}$ Technical University of Ko\v{s}ice, Ko\v{s}ice, Slovak Republic\\
$^{108}$ The Henryk Niewodniczanski Institute of Nuclear Physics, Polish Academy of Sciences, Cracow, Poland\\
$^{109}$ The University of Texas at Austin, Austin, Texas, United States\\
$^{110}$ Universidad Aut\'{o}noma de Sinaloa, Culiac\'{a}n, Mexico\\
$^{111}$ Universidade de S\~{a}o Paulo (USP), S\~{a}o Paulo, Brazil\\
$^{112}$ Universidade Estadual de Campinas (UNICAMP), Campinas, Brazil\\
$^{113}$ Universidade Federal do ABC, Santo Andre, Brazil\\
$^{114}$ Universitatea Nationala de Stiinta si Tehnologie Politehnica Bucuresti, Bucharest, Romania\\
$^{115}$ University of Cape Town, Cape Town, South Africa\\
$^{116}$ University of Derby, Derby, United Kingdom\\
$^{117}$ University of Houston, Houston, Texas, United States\\
$^{118}$ University of Jyv\"{a}skyl\"{a}, Jyv\"{a}skyl\"{a}, Finland\\
$^{119}$ University of Kansas, Lawrence, Kansas, United States\\
$^{120}$ University of Liverpool, Liverpool, United Kingdom\\
$^{121}$ University of Science and Technology of China, Hefei, China\\
$^{122}$ University of South-Eastern Norway, Kongsberg, Norway\\
$^{123}$ University of Tennessee, Knoxville, Tennessee, United States\\
$^{124}$ University of the Witwatersrand, Johannesburg, South Africa\\
$^{125}$ University of Tokyo, Tokyo, Japan\\
$^{126}$ University of Tsukuba, Tsukuba, Japan\\
$^{127}$ Universit\"{a}t M\"{u}nster, Institut f\"{u}r Kernphysik, M\"{u}nster, Germany\\
$^{128}$ Universit\'{e} Clermont Auvergne, CNRS/IN2P3, LPC, Clermont-Ferrand, France\\
$^{129}$ Universit\'{e} de Lyon, CNRS/IN2P3, Institut de Physique des 2 Infinis de Lyon, Lyon, France\\
$^{130}$ Universit\'{e} de Strasbourg, CNRS, IPHC UMR 7178, F-67000 Strasbourg, France, Strasbourg, France\\
$^{131}$ Universit\'{e} Paris-Saclay, Centre d'Etudes de Saclay (CEA), IRFU, D\'{e}partment de Physique Nucl\'{e}aire (DPhN), Saclay, France\\
$^{132}$ Universit\'{e}  Paris-Saclay, CNRS/IN2P3, IJCLab, Orsay, France\\
$^{133}$ Universit\`{a} degli Studi di Foggia, Foggia, Italy\\
$^{134}$ Universit\`{a} del Piemonte Orientale, Vercelli, Italy\\
$^{135}$ Universit\`{a} di Brescia, Brescia, Italy\\
$^{136}$ Variable Energy Cyclotron Centre, Homi Bhabha National Institute, Kolkata, India\\
$^{137}$ Warsaw University of Technology, Warsaw, Poland\\
$^{138}$ Wayne State University, Detroit, Michigan, United States\\
$^{139}$ Yale University, New Haven, Connecticut, United States\\
$^{140}$ Yonsei University, Seoul, Republic of Korea\\
$^{141}$  Zentrum  f\"{u}r Technologie und Transfer (ZTT), Worms, Germany\\
$^{142}$ Affiliated with an institute covered by a cooperation agreement with CERN\\
$^{143}$ Affiliated with an international laboratory covered by a cooperation agreement with CERN.\\

\end{flushleft} 